\documentclass
{aastex701}

\usepackage{graphicx}
\usepackage{lipsum}
\usepackage{soul}

\usepackage{amsmath}
\usepackage[caption=false]{subfig}
\usepackage{multirow}
\usepackage{comment}

\newcommand{\beq}{\begin{equation}}
\newcommand{\eeq}{\end{equation}}
\newcommand{\be}{\begin{eqnarray}}
\newcommand{\ee}{\end{eqnarray}}

\def\d{\partial}
\def\+{\dagger}

\def\<{\langle}
\def\>{\rangle}

\newcommand{\Lqcd}{\Lambda_{\mathrm{QCD}}}

\begin{document}

\title{Evolving Dark Energy Is Vacuum Energy After All}

\author[orcid=0009-0000-2992-3157,sname='Dong Ha Lee']{Dong Ha Lee}
\affiliation{School of Mathematical and Physical Sciences, University of Sheffield, UK}
\email[show]{dhlee1@sheffield.ac.uk}

\author[orcid=0000-0001-8311-8941,sname='Carsten van de Bruck']{Carsten van de Bruck}
\affiliation{School of Mathematical and Physical Sciences, University of Sheffield, UK}
\email[show]{c.vandebruck@sheffield.ac.uk}

\author[orcid=0000-0001-8408-6961,sname='Eleonora Di Valentino']{Eleonora Di Valentino}
\affiliation{School of Mathematical and Physical Sciences, University of Sheffield, UK}
\email[show]{e.divalentino@sheffield.ac.uk}

\author[orcid=0000-0002-2637-8728]{Ludovic Van Waerbeke}
\affiliation{Department of Physics and  Astronomy, University of British Columbia, Vancouver, B.C. V6T 1Z1, Canada}
\email[show]{waerbeke@phas.ubc.ca}  

\author[orcid=0000-0002-2049-228X]{Ariel Zhitnitsky} 
\affiliation{Department of Physics and  Astronomy, University of British Columbia, Vancouver, B.C. V6T 1Z1, Canada}
\email[show]{arz@phas.ubc.ca}

\begin{abstract}
We investigate a physically motivated model of dynamical dark energy arising from the non-perturbative topological structure of the Quantum Chromodynamics (QCD) vacuum.
The model introduces no new fundamental field or propagating degree of freedom: the dark energy (DE) density emerges as a global vacuum response to an expanding spacetime.
We develop the first comprehensive cosmological implementation of this QCD-DE scenario and confront it with current observations, including Planck, ACT and SPT-3G cosmic microwave background data, DESI DR2 baryon acoustic oscillation measurements, and Type Ia supernova samples from Pantheon+ and DES-Dovekie.
We compare the model with $\Lambda\mathrm{CDM}$ and $w_0w_a\mathrm{CDM}$ cosmologies.
The model provides an excellent fit to the data and reproduces the late-time DE evolution preferred by DESI.
The model naturally predicts effective phantom crossing behaviour at intermediate redshifts ($z\sim0.67$) while avoiding the instabilities associated with phantom scalar fields.
Using goodness-of-fit statistics and Bayesian model-selection tools, including Akaike and Deviance
Information Criteria and Bayesian evidence estimated from Markov-Chain Monte Carlo chains, we find that the QCD-induced model is consistently favoured over $\Lambda\mathrm{CDM}$ for the full combination of early and late-time datasets.
Unlike the conventional descriptions of dynamical DE, support for QCD-DE in Bayesian evidence remains more consistent across datasets, suggesting that a physically
motivated departure from a cosmological constant may provide a more economical description of the expansion history preferred by current observations.
\end{abstract}


\section{Introduction}\label{sec:introduction}

The standard cosmological model, $\Lambda$CDM, has been remarkably successful in describing a broad range of observations spanning the entire history of the Universe. With only six free parameters, it provides an excellent fit to measurements of the cosmic microwave background (CMB)~\citep{Planck:2018vyg,AtacamaCosmologyTelescope:2025blo,SPT-3G:2025bzu}, large-scale structure~\citep{Wright:2025xka,DES:2026fyc}, baryon acoustic oscillations (BAO)~\citep{eBOSS:2020yzd,DESI:2025zgx}, and Type Ia supernovae (SNe Ia)~\citep{Riess:1998cb,Perlmutter:1998np,Brout:2022vxf}, establishing itself as the current concordance model of cosmology. Within this framework, the present-day accelerated expansion of the Universe is attributed to a cosmological constant $\Lambda$, corresponding to a vacuum energy density with equation of state $w=-1$.

Despite its empirical success, several challenges remain. On the observational side, persistent discrepancies between independent measurements of key cosmological parameters have raised questions about the completeness of the standard model~\citep{CosmoVerseNetwork:2025alb}. The most prominent example is the Hubble constant tension, namely the disagreement between early- and late-Universe determinations of the current expansion rate~\citep{DiValentino:2021izs,H0DN:2025lyy}. More recently, increasing attention has focused on the possibility that the dark energy sector itself may exhibit departures from a pure cosmological constant~\citep{DESI:2025zgx}. From a theoretical perspective, the cosmological constant suffers from long-standing conceptual difficulties, including the fine-tuning and coincidence problems, which have motivated extensive efforts to explore alternative explanations for cosmic acceleration.

A major development in this context has been provided by the second data release of the Dark Energy Spectroscopic Instrument (DESI)~\citep{DESI:2025zgx}. The unprecedented precision of the DESI BAO measurements, extending over the redshift range $0.1 \lesssim z \lesssim 4.2$, has enabled increasingly stringent tests of the late-time expansion history. When combined with contemporary supernova samples and CMB observations~\citep{Hoyt:2026fve,Rubin:2026qdt,Popovic:2025glk,DES:2025sig,DES:2026jmi}, these data have revealed a preference for departures from a cosmological constant in several dynamical dark energy parametrizations~\citep{Li:2026asg,Najafi:2024qzm,Giare:2024gpk,Giare:2024oil,Jiang:2024xnu,RoyChoudhury:2024wri,Giare:2025pzu,Kessler:2025kju,RoyChoudhury:2025dhe,Scherer:2025esj,Wolf:2025jlc,Santos:2025wiv,Specogna:2025guo,Cheng:2025lod,Cheng:2025hug,Li:2025vuh,Lee:2025pzo,Fazzari:2025lzd,Smith:2025icl,Herold:2025hkb,Cheng:2025yue,Gokcen:2026pkq,Ishak:2025cay,Najafi:2026kxs,Yang:2026yaq,Kessler:2026dbi,Montefalcone:2026iga}. In particular, analyses based on the Chevallier-Polarski-Linder (CPL) \citep{Chevallier:2000qy,Linder:2002et} equation of state and related extensions frequently find indications that the dark energy equation of state evolves with redshift and may cross the phantom divide during cosmic history~\citep{Ozulker:2025ehg}.

At the same time, the interpretation of these results remains far from settled. The DESI collaboration quantified the preference for dynamical dark energy primarily through improvements in goodness-of-fit relative to $\Lambda$CDM. Subsequent studies have confirmed that a variety of evolving dark energy models can provide statistically significant improvements over the standard cosmological constant description. However, model comparison based on Bayesian evidence often paints a different picture. While the data may favour departures from $\Lambda$CDM at the level of parameter estimation, the improvement is frequently insufficient to overcome the Occam penalty associated with the introduction of additional degrees of freedom. As a consequence, Bayesian model selection generally continues to prefer the simpler $\Lambda$CDM framework~\citep{Ong:2026tta}.

The possibility of a dynamical dark energy component has nevertheless stimulated an enormous theoretical effort. A wide range of models have been proposed, including quintessence scenarios based on canonical scalar fields (see e.g. ~\citealp{Wetterich:1987fm,Wetterich:1994bg,Caldwell:1997ii,Barreiro:1999zs}), phantom models characterized by equations of state below $w=-1$ \citep{Caldwell:1999ew,Singh:2003vx,Dabrowski:2003jm}, k-essence constructions with non-canonical kinetic terms~\citep{Armendariz-Picon:2000ulo,Malquarti:2003hn}, and early dark energy models designed to modify the pre-recombination expansion history (see e.g. ~\citealp{Karwal:2016vyq,Poulin:2018cxd,Kamionkowski:2022pkx}). While these frameworks can reproduce a rich variety of cosmological behaviours, they often introduce additional theoretical challenges. Depending on the specific realization, such models may require finely tuned potentials, specific initial conditions, violations of energy conditions, or the introduction of new light degrees of freedom whose origin remains unclear within models of particle physics beyond the standard model. 

An alternative possibility is that the accelerated expansion is not driven by the dynamics of a new field at all. A number of approaches have explored this direction, including modified gravity theories~\citep{Carroll:2003wy,Chiba:2003ir,Flanagan:2003rb,Brookfield:2006mq}, running vacuum models~\citep{Sola:2016jky}, and other scenarios in which the effective dark energy density emerges from global properties of spacetime or vacuum structure rather than from propagating dynamical degrees of freedom~\citep{Benisty:2018qed,Banerjee:2019kgu}. Such frameworks offer the intriguing possibility of explaining cosmic acceleration without introducing new fundamental fields while potentially alleviating some of the conceptual difficulties associated with conventional dark energy models.

Recently, a particularly interesting realization of this idea was discussed in~\citealp{VanWaerbeke:2025shm}, building upon earlier works~\citep{Zhitnitsky:2013pna,Zhitnitsky:2015dia,Barvinsky:2017lfl}. In this framework, dark energy originates from a special contribution to the QCD vacuum energy associated with the global structure of the gauge field. The QCD gauge field admits distinct classes of vacuum configurations that cannot be continuously deformed into one another; these are commonly referred to as topologically distinct vacuum configurations.
Classically, these configurations are degenerate and have the same vacuum energy.
Quantum mechanically, tunnelling between them mixes the different configurations and shifts the energy of the true quantum vacuum.
A useful analogy is a quantum particle in a one-dimensional periodic potential: the different minima have the same classical energy, but quantum tunnelling between them shifts the ground-state energy of the system.
In this analogy with QCD, the horizontal axis does not represent a spatial coordinate, but rather a collective coordinate labelling the different QCD vacuum configurations.
Consequently, the tunnelling should not be pictured as the nucleation and expansion of bubbles occupying different regions of space; instead, it contributes to a single quantum vacuum state affecting the Universe globally.
The central proposal of QCD-DE is that this tunnelling contribution is sensitive to the large-scale geometry of spacetime and therefore differs between Minkowski and an expanding Universe.

An appealing feature of this proposal is that it naturally connects the observed dark energy scale to the fundamental QCD scale, $\Lambda_{\rm QCD}\sim 170,{\rm MeV}$. In this picture, the onset of cosmic acceleration is not associated with the emergence of a new ultra-light scalar field or a finely tuned vacuum energy, but instead arises from the response of the QCD vacuum to the global structure of an expanding Universe. The framework therefore offers a fundamentally different perspective on the nature of dark energy and on the origin of the observed late-time acceleration.

The purpose of this work is to investigate the cosmological implications of this QCD-induced dark energy scenario using current cosmological observations. In particular, we confront the model with the latest combination of CMB, DESI DR2 BAO, and Type Ia supernova data, and compare its performance against both $\Lambda$CDM and conventional dynamical dark energy parametrizations. Given the current tension between indications for evolving dark energy based on improvements in goodness-of-fit and Bayesian model-selection results that often continue to favour $\Lambda$CDM, this framework provides an especially interesting testing ground. It allows us to explore whether a physically motivated departure from a cosmological constant can reproduce the preferred late-time expansion history while remaining competitive under rigorous Bayesian model comparison.

This paper is organized as follows. In Section~\ref{sec:physicalDE} we review the theoretical framework underlying QCD-induced dark energy and describe its cosmological implementation. Section~\ref{sec:methodology} presents the datasets, methodology, and statistical tools used in our analysis. Our cosmological constraints and model-comparison results are discussed in Section~\ref{Sec:resuts}. Finally, Section~\ref{sec:concl} summarizes our conclusions and outlines future directions for this framework.

\section{A physical model of dark energy}
\label{sec:physicalDE}

\subsection{Generating Dark Energy from QCD topological sectors}\label{subsec:deSitter vacuum}

In a spatially flat universe, the Friedmann equation takes the form
\begin{equation}\label{eq:lcdm friedmann}
	H^2(z)=\frac{8\pi G}{3}\left[\rho_m(z)+\rho_r(z)+\rho_{\rm DE}(z)\right],
\end{equation}
where $(1+z)=a^{-1}$ is the redshift, $a$ is the scale factor, $H$ denotes the Hubble parameter, and $\rho_m$, $\rho_r$, and $\rho_{\rm DE}$ represent the energy densities of matter, radiation, and dark energy (DE), respectively.

In the approach of~\citealp{Zhitnitsky:2013pna,Zhitnitsky:2015dia,Barvinsky:2017lfl}, the DE energy density is written as the difference between vacuum energy densities, $\Delta\varepsilon_{\rm vac}$, given by
\begin{equation}
	\label{eq:casimir de}
	\rho_{\rm DE}\equiv\Delta\varepsilon_{\rm vac}
	=
	\varepsilon_{\rm FLRW}^{\rm vac}
	-
	\varepsilon_{\mathrm{Mink}}^{\rm vac},
\end{equation}
where $\varepsilon_{\rm FLRW}^{\rm vac}$ is the vacuum energy density in an expanding FLRW spacetime and $\varepsilon_{\mathrm{Mink}}^{\rm vac}$ is the vacuum energy density in Minkowski spacetime.

This definition of the vacuum energy density was first proposed by~\citealp{Zeldovich:1967gd}, who argued that
\begin{equation}
\Delta\varepsilon_{\rm vac}
=
\varepsilon_{\rm FLRW}^{\rm vac}
-
\varepsilon_{\mathrm{Mink}}^{\rm vac}
\sim G m_p^6,
\end{equation}
and therefore must be proportional to the gravitational constant $G$, with $m_p$ being the proton mass, a quantity set by the QCD scale. In the following decades, numerous studies adopting the same definition of $\rho_{\rm DE}$ in Einstein's field equations have appeared in the literature across several disciplines, including particle physics, cosmology, and condensed-matter physics (see \citealt{Zhitnitsky:2015dia} for references and further details).
\par
The central idea of this proposal is that $\Delta\varepsilon_{\rm vac}$ describes the topological vacuum energy of QCD. The effect originates from the non-local structure of the QCD vacuum and does not rely on the local dynamics of a new field. The non-perturbative vacuum structure of QCD contains multiple degenerate topological sectors which become slightly modified in a time-dependent background such as an expanding universe relative to Minkowski spacetime. As a result, the tunnelling\footnote{Tunnelling between vacuum states is a non-local topological effect driven by a global restructuring of the vacuum, but it does not violate causality.} between these sectors is argued to generate a small correction to the vacuum energy proportional to the Hubble parameter $H$. Consequently, $\varepsilon_{\rm FLRW}^{\rm vac}\neq\varepsilon_{\mathrm{Mink}}^{\rm vac}$, and the vacuum energy difference $\Delta\varepsilon_{\rm vac}$ defined in Eq.~(\ref{eq:casimir de}) receives a non-vanishing contribution proportional to $H$.
\par
At first sight, this appears counter-intuitive, since QCD describes short-range nuclear physics and should therefore be insensitive to cosmological scales. However, the effect does not arise from propagating degrees of freedom, but from the structure of the QCD vacuum itself. In particular, the vacuum energy contains a non-dispersive contribution (the contact term in the topological susceptibility) which is not controlled by the QCD mass gap and is therefore not restricted to microscopic distances. Consequently, this component can remain sensitive to arbitrarily large scales, including the cosmological expansion\footnote{This feature is unique to QCD in the standard model since the mass gap of the electroweak (EW) sector come from the Higgs mechanism. Therefore, vacuum energy is not sensitive to the topological features of the EW gauge fields.There is no analogue of the strongly coupled, confining QCD vacuum with a large IR-sensitive topological susceptibility. QCD has confinement, chiral symmetry breaking, the $\eta$' meson, and a large nonperturbative topological suceptibility. The weak sector has a Higgs phase and exponentially suppressed tunnelling at late times.}.
\par
Non-perturbative computation of the QCD vacuum $\Delta\varepsilon_{\rm vac}$ for arbitrary geometries is currently infeasible due to several technical challenges (see~\citealp{Zhitnitsky:2015dia} for a detailed discussion). Nevertheless, a calculation can be performed for Minkowski spacetime in $\mathbb{R}^3 \times S^1$, yielding\footnote{All numerical coefficients of order one in the computations can be absorbed by redefinition of a single dimensional parameter $\Lqcd$.}
\begin{equation}\label{eq:Minkowski vacuum}
	\varepsilon^{\rm vac}_{\rm Mink} = -\Lambda_{\rm QCD}^4,
\end{equation}
where $\Lambda_{\rm QCD} \sim 100 \, \mathrm{MeV}$ is the QCD energy scale. A similar calculation can be performed for the relativistic hyperbolic spacetime $\mathbb{H}^3_\kappa \times S^1$, as shown in~\citealp{Zhitnitsky:2015dia}, yielding
\begin{equation}\label{eq:hyperbolic vacuum}
	\varepsilon^{\rm vac}_{\rm hyperbolic} = -\Lambda_{\rm QCD}^4
	\left(1 - c_\kappa \frac{\kappa}{\Lambda_{\rm QCD}} \right),
\end{equation}
where $\kappa$ is the dimensional parameter characterizing the hyperbolic spacetime and $c_\kappa$ is a dimensionless numerical parameter of order $O(1)$. The vacuum energy density according to the prescription of Eq.~(\ref{eq:casimir de}) is therefore given by
\begin{equation}\label{eq:kappa energy}
	\Delta\varepsilon_{\rm vac}=c_\kappa\,\kappa\,\Lambda_{\rm QCD}^3.
\end{equation}
This result is mathematically exact and demonstrates that, in QCD, the global restructuring of the vacuum leads to a correction term linear in $\kappa$. This is in contrast to the naively expected $\kappa^2$ correction.
\par
The conjecture put forward by~\citealp{Zhitnitsky:2015dia} is that this result can be extended to an expanding universe in the de Sitter limit, where the departure from Minkowski spacetime is characterized by a constant Hubble parameter $\overline{H}$, through the replacement $c_\kappa \rightarrow c_H$ and $\kappa \rightarrow \overline{H}$, leading to
\begin{equation}\label{eq:deSitter vacuum}
	\varepsilon_{\rm de\,Sitter}^{\rm vac} = -\Lambda_{\rm QCD}^4
	\left(1 - c_H \frac{\overline{H}}{\Lambda_{\rm QCD}} \right).
\end{equation}
This conjecture is supported by the arguments presented in~\citealp{Zhitnitsky:2015dia,Barvinsky:2017lfl}, which state that the key element of the computations is the topological structure of QCD, rather than a specific feature of the geometry.
\par
This conjecture is further supported by a number of other computations. In particular, the linear dependence on an external parameter has been tested in a weakly coupled so-called ``deformed QCD'' model, where all computations are under complete theoretical control. In that model, the sensitivity of the vacuum energy to very large distances can be studied by enclosing the system in a box of size $\mathbb{L}$. As shown in~\citealp{Thomas:2012ib}, the corrections to the vacuum energy scale linearly with the inverse size, $\sim \mathbb{L}^{-1}$, a behaviour analogous to the role played by the parameter $\kappa$ in Eq.~(\ref{eq:hyperbolic vacuum}).
\par
Equation~(\ref{eq:deSitter vacuum}) is also consistent with lattice simulations presented in~\citealp{Yamamoto:2014vda}. Computations of the vacuum energy in a background characterized by $H$ represent a very challenging technical problem. Indeed, one must use an imaginary $H$ to make the path integral suitable for lattice computations. Instead of computing the vacuum energy directly (which is a very difficult task on the lattice), the author investigates a different observable, namely the rate of particle production in a de Sitter background with constant $H$. The results show that the production rate is linearly proportional to the Hubble constant, scaling as $\sim H$, rather than the naively expected $H^2$. This provides support for the expectation that the vacuum energy, if computed on the lattice, would also exhibit a linear scaling with $H$.
\par
One can explicitly see why the standard arguments based on locality (which suggest a $\kappa^2$ correction) are strongly violated by tunnelling events. As shown in the computations of~\citealp{Zhitnitsky:2015dia}, the linear correction $\sim \kappa$ is explicitly proportional to a global gauge-invariant observable that cannot be reduced to the local curvature. In other words, this correction arises from non-local configurations and cannot be expressed solely in terms of the local curvature, which would instead lead to a quadratic correction proportional to $\kappa^2$.
\par
The assumption (\ref{eq:deSitter vacuum}) immediately leads to the effective DE density:
\begin{equation}\label{eq:deSitter energy}
    \overline{\rho_\mathrm{DE}}
	\equiv
	\varepsilon_{\rm de\,Sitter}^{\rm vac}
	-
	\varepsilon_{\rm Mink}^{\rm vac}
	=
	c_H \,\overline{H}\,\Lambda_{\rm QCD}^3.
\end{equation}
In the asymptotic de Sitter regime, the Hubble constant $\overline{H}$ and the energy density can be expressed as:
\begin{equation}\label{eq:deSitter Hubble}
    \overline{H}=c_H\frac{8\pi G\Lqcd^3}{3}, ~~  \overline{\rho_{\rm DE}}=c_H^2 \frac{8\pi G\Lqcd^6}{3}
\end{equation}
by substituting the expression for the DE energy density (\ref{eq:deSitter energy}) into the Friedmann equation (\ref{eq:lcdm friedmann}). This corresponds to the estimate originally proposed by~\citealp{Zeldovich:1967gd}, provided one replaces $m_p\rightarrow \Lqcd$ in his expression.
\par
A back-of-the-envelope calculation with $\Lqcd\sim 100~ \rm MeV$ and $c_H \propto m_q/\Lqcd\sim (0.01-0.02)$, which incorporates the conventional suppression factor $\propto m_q$ associated with the topological nature of the tunnelling process, leads to:
\be
\label{eq:order of magnitude}
\overline{H}&\sim&(0.56-1.12)\cdot 10^{-33}~{\rm eV}, \nonumber\\
&\sim&(26-53)~{\rm km/s/Mpc},\\
\overline{\rho_{\rm DE}} &\sim&\left[(1.5-2.2)\cdot 10^{-3}~{\rm eV}\right]^4, \nonumber
\ee
which are of the same order of magnitude as the values observed today. This represents a highly non-trivial aspect of the proposal, as all quantities, $\overline{H}$ and $\rho_{\rm DE}$, are governed by a single dimensional parameter, $\Lqcd$. It also answers the question ``why is DE relevant now?'', i.e. where the scale $\sim 10^{-3}\,\rm eV$ comes from, which does not correspond to any particular scale in particle physics.
\par
The direct consequence of Eq.~(\ref{eq:deSitter energy}) is that all effects discussed here are global in nature and cannot be formulated in terms of any local effective quantum field theory (QFT). This is fundamentally different from the conventional treatment where dark energy is described as a new scalar field $\phi$ (``quintessence'', ``k-essence'', ``phantom fields'', etc.) and an effective potential $V(\phi)$. These approaches may be somewhat restricted by issues such as the fine-tuning problem, instabilities, violations of unitarity, and other fundamental principles of QFT. However, by assuming a QCD-induced DE mechanism, we may be able to introduce a description of an effective DE without new dynamical degrees of freedom. As a result, our framework is free from any violations of the fundamental principles of QFT mentioned above.

As emphasized above, this form of DE does not involve any propagating degrees of freedom.
It is intrinsically topological in nature, and it is therefore not physically meaningful to assign dynamical or thermodynamical attributes to it that would ordinarily characterize a fluid, such as pressure.
Nevertheless, one may formally derive what the pressure and energy density would be if the topological nature of this DE were disregarded:
\be
\label{eq:deSitter pressure}
\overline{P_{\rm DE}} = -  c_H \Lqcd^3 \overline{H}, ~~~
\overline{\rho_{\rm DE}}=  +   c_H \Lqcd^3 \overline{H}.
\ee
This leads to the effective equation of state parameter as inferred by an observer:
\be
\label{eq:deSitter EoS}
\overline{w} &=& \frac{\overline{P_{\rm DE}}}{\overline{\rho_{\rm DE}}}= -1
\ee
which is the expectation for a pure de Sitter universe.
\par
We conclude this overview subsection on the QCD-DE proposal with the following comment. Can we test some of these key ideas on the nature of dark energy using a tabletop experiment?
\par
The ultimate answer is affirmative: yes, as it represents a genuine novel physical phenomenon rather than a mere formal manipulation of the equations.
The basic concept behind such an experiment is to detect a novel contribution to the Casimir vacuum energy in Maxwell theory, as proposed in~\citealp{Cao:2013na,Cao:2015uza,Yao:2016bps}.
This contribution, known as the Topological Casimir Effect (TCE), arises from non-dispersive contributions to the vacuum energy expressed in terms of tunnelling events rather than from conventional quantum fluctuations of propagating photons with two physical transverse polarizations (the dispersive contribution). Although this correction to the Casimir pressure has not yet been measured, its detection would provide direct empirical support for the topological origin of dark energy. Therefore, in essence, the nature of the vacuum DE advocated in this work is a Casimir-type energy analogous to the TCE, which may (potentially) be measured in a laboratory.

\subsection{Before the de Sitter limit with a time dependent $H(t)$}\label{subsec:adiabatic approximation}

Unlike the discussion above, our universe has yet to reach a pure de Sitter regime with a constant $\overline{H}$ and has undergone transitions from radiation domination to matter domination and finally to the late-time dark energy domination observed today. To use the QCD vacuum description of DE in the present epoch,~\citealp{VanWaerbeke:2025shm} replace $\overline{H}$ in Eq.~(\ref{eq:deSitter energy}) with the time-dependent Hubble parameter $H(t)$ to capture the behaviour as the universe approaches the de Sitter limit.
\par
A justification for this modification can be found in~\citealp{Zhitnitsky:2015dia}, whose final result is given by Eq.~(\ref{eq:hyperbolic vacuum}).
The key element of this analysis is the correction $(\kappa/\Lqcd)$, which originates from the zero-mode determinant appearing in the tunnelling transition rate. Such computations inherently involve integration over collective variables (spatial and temporal coordinates) extending to distances of order $\kappa^{-1}$.
\par
However, if a time-dependent perturbation describing a deviation from pure de Sitter space with constant $\overline{H}$ is introduced on a scale $\omega \gg \kappa$, where $\omega\sim {|\dot{H}|}/{H}$ denotes the characteristic frequency of the perturbation, then, for $\kappa\rightarrow \overline{H}$, a suppression factor $\propto (\kappa/\omega) \ll 1$ arises. This occurs because the zero mode ceases to be an exact zero mode, and the associated collective variable is correspondingly reduced by a factor $\propto (\kappa/\omega)$. Consequently, the integration can no longer extend to its maximal range $\kappa^{-1}$, but is instead effectively truncated at the smaller scale $\omega^{-1}$, reflecting the strong sensitivity of the zero mode to large distances (see \citealp{VanWaerbeke:2025shm} for further discussions on the justifications of the adiabatic approximation).
\par
If we choose to extend the definition of the linearly dependent DE density to a time-dependent Hubble parameter, $\rho_{\rm DE}(t)\approx c_H \Lqcd^3 H(t)$, it would appear, from the considerations above, that this modification is justified only if the suppression is small, i.e. $\omega \ll \kappa$. We can express this condition as the adiabatic approximation
\be
\label{eq:adiabatic approximation}
\frac{|\dot{H}|}{H} \ll \overline{H},
\ee
which is required for $\rho_{\rm DE}(t)\propto H(t)$. Otherwise, we would expect $\rho_{\rm DE}$ to experience additional suppression, as described above. This suppression may be interpreted as a reduction in the numerical coefficient $c_H$ when the adiabatic approximation is not satisfied. While, in principle, it may be possible to compute the corrections to $c_H$ for a given $H(t)$, this is not currently feasible.
\par
The adiabatic condition can be recast in terms of the total equation of state $w_\mathrm{tot}\equiv p_\mathrm{tot}/\rho_\mathrm{tot}$, giving the condition
\begin{equation}\label{eq:adiabatic total eos}
    \frac{3}{2}\frac{H}{\overline{H}}\vert1+w_\mathrm{tot}\vert\ll1.
\end{equation}
The condition above implies that, for the early universe, where $H\gg\overline{H}$ and $w_\mathrm{tot}=1/3$ (RD) or $w_\mathrm{tot}=0$ (MD), QCD-DE is a subdominant effect. Only at late times, when $H\sim\overline{H}$ and sufficient QCD-DE generation has occurred to satisfy the adiabatic condition, does the DE density evolve as $\rho_\mathrm{DE}(t)\approx c_H\Lqcd^3H(t)$ and asymptotically approach Eq.~(\ref{eq:deSitter energy}) in the de Sitter limit.
\par
For the early universe, we are therefore required to impose an additional suppression which would depend on $\dot{H}/H$ in a highly non-trivial manner. While it is not possible for us to compute this suppression exactly at an arbitrary time, we expect $H(t)$ to be a monotonically decreasing function of $t$ and, at the present epoch, we expect the universe to at least marginally satisfy condition~(\ref{eq:adiabatic approximation}) such that we have a DE-dominated epoch with $w_\mathrm{tot}<-1/3$.
\par
To impose this additional suppression and deviation from $\rho_\mathrm{DE}\propto H$ at early times, we introduce a monotonic time-dependent function $\beta(z)$ such that
\begin{equation}\label{eq:vacuum density}
    \rho_\mathrm{DE}(z) = [\beta(z)\cdot c_H]\,\Lqcd^3H(z), \qquad \beta(z)\in (0,1).
\end{equation}
This switch function $\beta(z)$ must tend to one as $z\rightarrow -1$ (de Sitter limit) such that we recover the result in Eq.~(\ref{eq:deSitter energy}). Also, for sufficiently large redshift ($z\rightarrow\infty$), we would expect DE to play a subdominant role relative to other components such as matter or radiation due to this suppression, such that $\beta(z\rightarrow\infty)\rightarrow0$.

The function $\beta(z)$ may be interpreted as a switch mechanism that activates dark energy at a given redshift. Within this framework, it provides a physically motivated parametrization of the dark energy component.\footnote{It is interesting to note that, on the phenomenological level, the ``switch'' feature has previously been introduced within the so-called critically emergent dark energy model (CEDE)~\citep{Banihashemi:2020wtb,Najafi:2026kxs}, where DE switches on at some redshift $z_c$. If we attempt to map this QCD-DE phenomenology onto the CEDE model formulated in relativistically covariant terms of some canonical scalar field $\phi$ with potential $V(\phi)$, it would lead to numerous fundamental problems discussed above, such as violations of unitarity, causality, internal instabilities, and related QFT problems, including a negative squared speed of sound, etc.; see e.g. the review~\citep{Cai:2009zp}.}
\par
It is important to emphasize that introducing a time-dependent $\beta(z)$ is not an ad hoc addition to the proposal. Rather, it serves as an effective phenomenological tool to account for the suppression factor $\propto (\kappa/\omega)$ discussed at the beginning of this subsection, which must be implemented in one way or another. The insertion of the $\beta(z)$ function in Eq.~(\ref{eq:vacuum density}) is precisely a specific implementation of this suppression when the universe deviates from a pure de Sitter state.
\par
From the model-independent constraints on the deceleration parameter $q(z)=-\left(1+\dot{H}/H^2\right)$ obtained by~\citealp{Fazzari:2025lzd}, we see that at a redshift of around $1$, $q(z\sim1)\sim0$, while at the present day, $q(z\sim0)\sim-0.4$. These correspond to $w_\mathrm{tot}(z\sim1)\sim-1/3$ and $w_\mathrm{tot}(z\sim0)\sim-0.6$, respectively. We can combine these numbers with the estimates for $\overline{H}$ from Eq.~(\ref{eq:order of magnitude}), which imply that $\mathcal{O}(H/\overline{H})\sim1$ at these redshifts, and we see that the condition (\ref{eq:adiabatic total eos}) may be marginally satisfied as the universe enters the present-day epoch.
\par
In the following sections, we describe the cosmology implied by this DE model and the parametrization of the switch function $\beta(z)$ used in our cosmological analysis.

\subsection{Friedmann equation with QCD-DE proposal (\ref{eq:vacuum density})}\label{subsec:friedmann with QCD}

Introducing the contribution of QCD-DE to the Friedmann equation (\ref{eq:lcdm friedmann}) leads to
\begin{equation}\label{eq:friedmann with DE}
	H^2(z)=\frac{8\pi G}{3}\rho_t(z)+\overline{H}H(z)\beta(z),
\end{equation}
where $\rho_t = \sum\rho_i$ is the total energy density of the components contributing to the stress-energy tensor ($i\in\{b,c,\gamma,\nu\}$), and we have recast Eq.~(\ref{eq:vacuum density}) in terms of the de Sitter Hubble constant $\overline{H}$ defined in Eq.~(\ref{eq:deSitter Hubble}). We can rearrange this equation to write the Hubble parameter explicitly as
\begin{equation}\label{eq:new Hubble}
	H(z)=\frac{\overline{H}\beta(z)}{2}+\sqrt{\left(\frac{\overline{H}\beta(z)}{2}\right)^2+\frac{8\pi G}{3}\rho_t(z)}.
\end{equation}
We note that if $\beta\rightarrow0$, we recover the standard expression $H=\sqrt{\frac{8\pi G}{3}\rho_t}$, while if $z\rightarrow-1$, implying that $\beta\rightarrow1$ and $\rho_t\rightarrow0$, we find that $H\rightarrow\overline{H}$, as required.
\par
The second Friedmann equation, $\dot{H} = -4\pi G(\rho_t+P_t)$, will also receive contributions from the QCD vacuum DE, such that
\begin{equation}\label{eq:second friedmann}
	2\dot{H}+3H^2 = -8\pi G(P_t+P_\mathrm{DE}).
\end{equation}
where $P_t$ is the total pressure of the components contributing to the stress-energy tensor and $P_\mathrm{DE}$ is the effective pressure contribution from the QCD-DE term.
\par
In similar fashion, we can write the effective energy density of the QCD-DE as
\begin{equation}\label{eq:DE density}
	\rho_\mathrm{DE} = \frac{3\overline{H}H\beta}{8\pi G}
\end{equation}
which preserves the standard form of the Friedmann equations.
\par
To complete the set of equations, we can differentiate Eq.~(\ref{eq:new Hubble}) and substitute the continuity equation $\dot{\rho}_t+3H(\rho_t+p_t)=0$ to obtain
\begin{equation}\label{eq:new Hdot}
	\dot{H} = \frac{H^2}{2H^2-\overline{H}H\beta}\left[\overline{H}H\frac{\mathrm{d}\beta}{\mathrm{d}\ln a}-8\pi G(\rho_t+P_t)\right].
\end{equation}
When $\beta\rightarrow0$, we recover the standard expression $\dot{H} = -4\pi G(\rho_t+P_t)$, and when $a\rightarrow\infty$, provided that $\frac{\mathrm{d}\beta}{\mathrm{d}\ln a}\rightarrow0$, we find that $\dot{H}\rightarrow0$, as required for a de Sitter geometry.
\par
A fluid description of dynamical dark energy typically parametrizes the dynamics with an equation of state, such as the CPL parametrization, which can be interpreted as a phenomenological description of a variety of scalar field models and is described by
\begin{equation}\label{eq:CPL}
	w(a) = w_0 + w_a(1-a)
\end{equation}
where $w_0$ is the value of the equation of state parameter today and $w_a = -\frac{\mathrm{d}w}{\mathrm{d}a}\vert_{a=1}$. If we choose to make direct comparisons with such models, we could rearrange Eq.~(\ref{eq:second friedmann}) and use the expression for $\dot{H}$ in Eq.~(\ref{eq:new Hdot}) to compute the effective equation of state ($w_\mathrm{DE}\equiv P_\mathrm{DE}/\rho_\mathrm{DE}$) of this model. It is worth noting that, unlike parametrizations of the equation of state, here the equation of state is merely something one can examine for a more ``apples-to-apples'' comparison with other dynamical dark energy models. The model itself is not dependent on the form of $w$ and therefore the value of $P_\mathrm{DE}$.
\par
Another differentiating feature of this model is that it is purely a background effect, since the deviation from $\Lambda\mathrm{CDM}$ is a non local phenomenon. This contrasts with other dynamical dark energy models, where dark energy corresponds to a physical degree of freedom and therefore introduces perturbations whenever it differs from a cosmological constant ($w=-1$).

\subsection{The switch function $\beta$}\label{subsec:beta}

In Section~\ref{subsec:adiabatic approximation}, we touched on the necessity for an additional time-dependent function $\beta(t)\in(0,1)$ to parameterize the suppression of QCD-DE at early times (corresponding to the universe not satisfying the adiabatic approximation (\ref{eq:adiabatic approximation})). In this section, we examine the form of $\beta$ itself and the choice of parametrizations of $\beta$ used in this paper.
\par
We note that for early-time radiation domination (RD), we require that in Eq.~(\ref{eq:new Hubble}), $\beta_\mathrm{RD}\rightarrow0$ in order to recover the standard RD Friedmann equation (i.e. DE is a subdominant component of the universe). In the de Sitter limit, we expect $\beta_\mathrm{dS}\rightarrow1$ by definition. If we assume that DE only dominates at late times ($z\sim1$), we would expect $\beta\sim0$ throughout most of the universe's history, only acquiring significance around some threshold redshift $z_\mathrm{q}\sim1$.
\par
We must also consider the derivative of $\beta$, which appears in the equation for $\dot{H}$, Eq.~(\ref{eq:new Hdot}), as $\frac{\mathrm{d}\beta}{\mathrm{d}\ln a}$. If QCD-DE is subdominant in the early universe, we require that the term $H\frac{\mathrm{d}\beta}{\mathrm{d}\ln a}\rightarrow0$, which recovers the standard expression for $\dot{H}$. As the energy densities of the stress-energy components are diluted by the expansion, we eventually expect the derivative term to dominate over the stress-energy terms. However, once this occurs, we then require that $\frac{\mathrm{d}\beta}{\mathrm{d}\ln a}$ decreases in order to satisfy the adiabatic approximation (\ref{eq:adiabatic approximation}) and eventually tends to $0$ in the de Sitter limit.
\par
In this paper, we consider a monotonic $\beta$ function parametrized by two parameters, $\{z_q,\Delta z_q\}$, as follows:
\begin{equation}
    \beta(z)\equiv\frac{A_q}{1+\exp\left(\frac{z-z_q}{\Delta z_q}\right)}\label{eq:beta exp}
\end{equation}

\begin{figure}
	\centering
	\hspace*{-0.5cm}\includegraphics[width=0.5\textwidth]{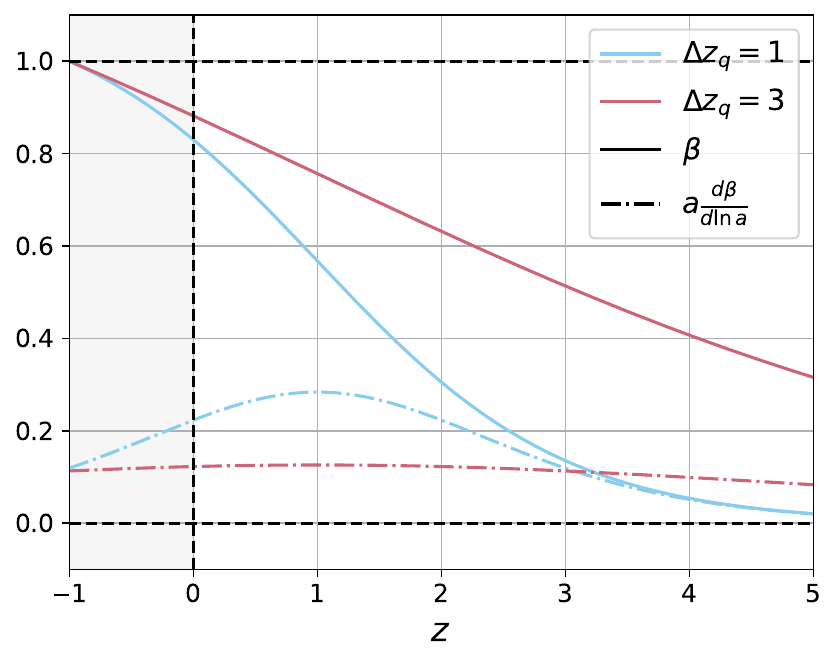}
	\caption{
    $\beta$ (solid) and $a\frac{\mathrm{d}\beta}{\mathrm{d}\ln a}$ (dashed) for $z_q=1$ and $\Delta z_q=1$ (blue) or $3$ (red). Changing $z_q$ would be equivalent to horizontal translation of the function in redshift space.}
	\label{fig:beta}
\end{figure}

The normalization factor $A_q$ is dependent on the choice of $z_q$ and $\Delta z_q$, and it is fixed such that $\beta(z=-1)=1$, where $z=-1$ corresponds to the de Sitter limit $a\rightarrow\infty$. The parameter $\Delta z_q$ represents the ``width'' of the transition into DE domination in redshift space. The choice of $\beta$ functions allows us to analytically compute its derivative:
\begin{equation}\label{eq:adbeta/dlna}
    a\frac{\mathrm{d}\beta}{\mathrm{d}\ln a} = \frac{\beta(A_q-\beta)}{A_q\Delta z_q}
\end{equation}
The forms of the switch function and its derivative are shown in Fig.~(\ref{fig:beta}). The plots show $a\frac{\mathrm{d}\beta}{\mathrm{d}\ln a}$, which is more convenient for computations, and since this quantity is finite at $z=-1$, the condition $\frac{\mathrm{d}\beta}{\mathrm{d}\ln a}\vert_\mathrm{dS}\rightarrow0$ is satisfied.

\subsection{QCD-DE Cosmology}\label{subsec:cosmology}

Putting the modifications to cosmology from Section~\ref{subsec:friedmann with QCD} and the parametrization of the QCD vacuum suppression from Section~\ref{subsec:beta} together, we obtain a complete description of the cosmology described by the QCD-DE model. The cosmology is an eight-parameter ($\Lambda\mathrm{CDM}+\{z_q,\Delta z_q\}$) model which is concordant with $\Lambda\mathrm{CDM}$ at early times, but deviates from both $\Lambda\mathrm{CDM}$ and fluid dynamical dark energy models as $\beta$ becomes significant.
\begin{figure*}
    \centering
    \subfloat[
    Form of $H(z)$ for different cosmological models. The value of $\overline{H}$ is obtained by matching the cosmology to the required present-day value $\Omega_{\mathrm{DE},0}=0.7$. All QCD-DE models with any $\beta(z)\in(0,1)$ predict a pure de Sitter behaviour with constant $\overline{H}$ in the far future ($z=-1$).%
    \label{fig:H(z) models}]{
    \includegraphics[width=0.48\textwidth]{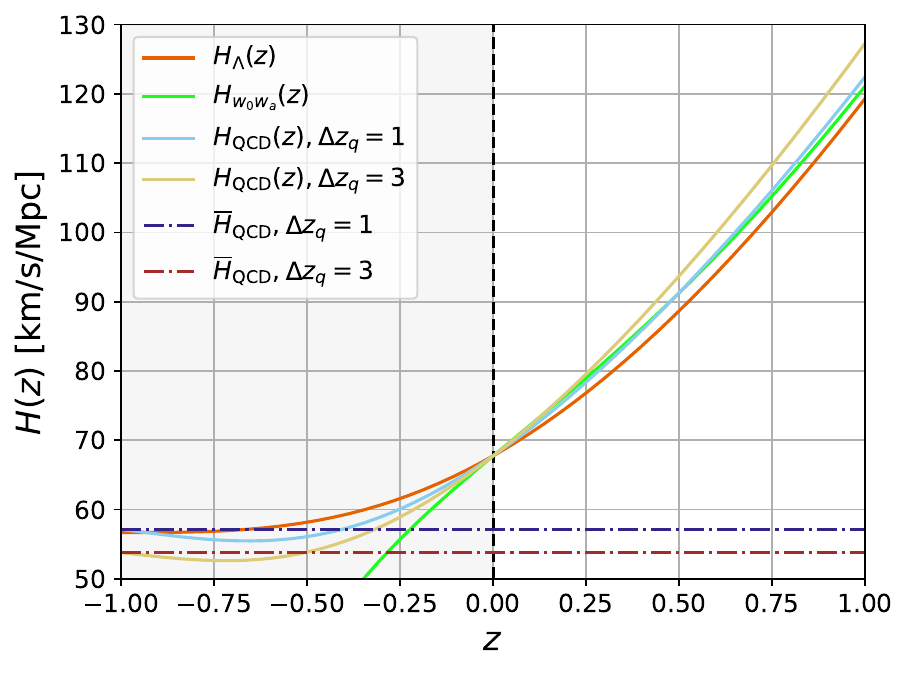}
    }
    \hfill
    \subfloat[The adiabatic condition $\vert\dot{H}\vert\,/\,\left(\overline{H}H\right)$ for $\Lambda$CDM and the QCD dark energy case. For $\Lambda\mathrm{CDM}$ we have defined $\overline{H}_{\Lambda{\rm CDM}}\equiv H_0\sqrt{\Omega_\Lambda}$. The region satisfying condition~(\ref{eq:adiabatic approximation}) is shaded green.
    \label{fig:adiabatic}]{
    \includegraphics[width=0.48\textwidth]{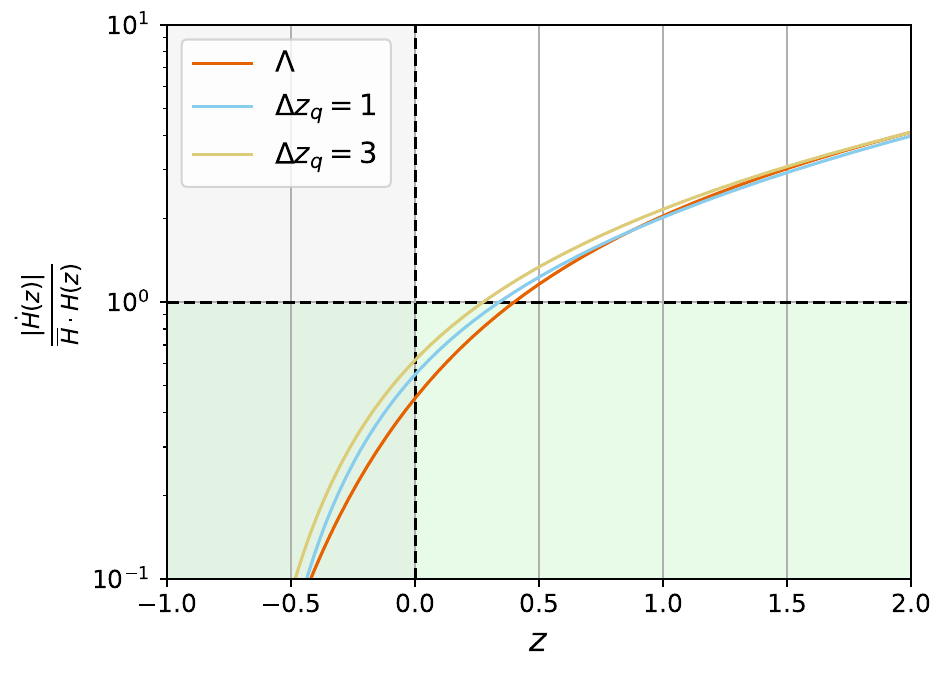}
    }
    \caption{
    Background evolution functions for $\Lambda\mathrm{CDM}$, $w_0w_a\mathrm{CDM}$, and the QCD-DE model ($\Delta z_q = 1$ or $3$) for a given set of cosmological parameters. The chosen cosmological parameters are as follows: $\Omega_m=0.3$, $H_0=67.75$; $w_0=-0.8$, $w_a=-0.5$ for $w_0w_a\mathrm{CDM}$; and $z_q=1$ for QCD-DE.}
    \label{fig:background cosmology}
\end{figure*}
\par
One can observe the differences in the cosmological evolution between the QCD-DE models and both $\Lambda\mathrm{CDM}$ and $w_0w_a\mathrm{CDM}$, an extension of $\Lambda\mathrm{CDM}$ using the CPL parametrization (\ref{eq:CPL}) to describe DE, in Fig.~(\ref{fig:H(z) models}). This figure demonstrates that the QCD-DE model is able to maintain physical consistency while having a dynamical DE for $z<0$, unlike the $w_0w_a\mathrm{CDM}$ model. Fig.~(\ref{fig:adiabatic}) also demonstrates how our choice of $\beta$ functions correctly parameterize the additional suppression of QCD-DE such that the adiabatic condition (\ref{eq:adiabatic total eos}) is only satisfied at late-times, corresponding to the epoch of DE domination.
\par
\begin{figure}
    \centering
    \includegraphics[width=0.5\linewidth]{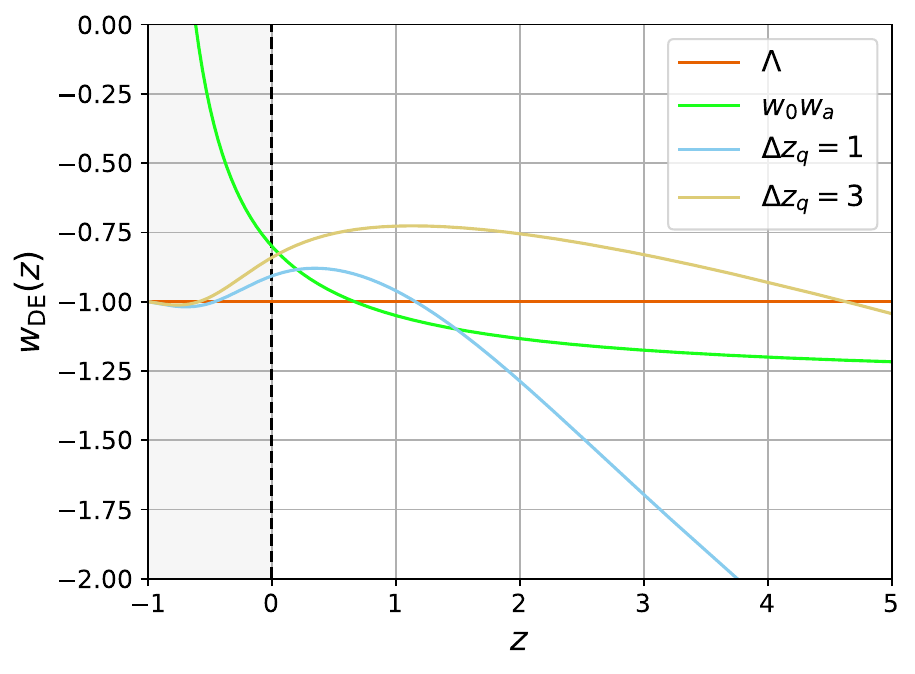}
    \caption{
    The equation of state of DE, $w_\mathrm{DE}\equiv P_\mathrm{DE}/\rho_\mathrm{DE}$, for $\Lambda\mathrm{CDM}$, $w_0w_a\mathrm{CDM}$, and the QCD-DE model ($\Delta z_q = 1$ or $3$).}
    \label{fig:w models}
\end{figure}

The equation of state of DE for the different models are shown in Fig.~(\ref{fig:w models}), where the effective equation of state of QCD-DE approaches $-1$ in the de Sitter regime, is quintessence-like today ($w_\mathrm{DE}>-1$), but was phantom in the past ($w_\mathrm{DE}<-1$). The phantom crossing in our model is consistent with the theory since QCD-DE is not a physical propagating field, but rather a global correction to the vacuum energy density due to the spacetime curvature, by the definition given in Eq.~(\ref{eq:casimir de}).
We also note that in the QCD-DE model, $w_\mathrm{DE}$ and consequently $P_\mathrm{DE}$ are not the subject of an analysis of the stability of the system, which would represent the conventional problem for any other model describing phantom behaviour; see additional comments and references in~\citealp{VanWaerbeke:2025shm}. This is because in the QCD-DE framework there are no propagating degrees of freedom, no additional scalar field $\phi$, no potential $V(\phi)$, and no canonical momentum corresponding to such a field.
\par
We modified the Cosmic Linear Anisotropy Solving System (\texttt{CLASS}) developed by~\citealp{Blas:2011rf} to compute the QCD-DE cosmologies for $\beta$ parametrized by Eq.~(\ref{eq:beta exp}).\footnote{\url{https://github.com/dlehdgk/class_qcd.git}} The cosmology described by the model has eight parameters and uses the constraint $1 = \sum\Omega_i$ to derive $\overline{H}$ given the choice of parameters and the value of $\Omega_\mathrm{DE}$ from the constraint equation. $A_q$, for a given choice of $z_q$ and $\Delta z_q$, is determined by inverting Eq.~(\ref{eq:beta exp}) with $z=-1$ and $\beta=1$. The resulting model has the same number of parameters as the popular $w_0w_a\mathrm{CDM}$ model, but describes a cosmology that is distinct from both $w_0w_a\mathrm{CDM}$ and $\Lambda\mathrm{CDM}$.
\par
Before concluding this section, we emphasize that this model, given the choice of parametrizations of $\beta$ adopted in this paper, is not an extension of $\Lambda\mathrm{CDM}$ in the sense that $\Lambda\mathrm{CDM}$ is not a nested model of QCD-DE.

\section{Methodology}\label{sec:methodology}

\subsection{Cosmological Datasets}\label{subsec:datasets}

The analysis in this paper involves several different combinations of datasets from a range of cosmological observations. The datasets can be classified into three groups: measurements of the Cosmic Microwave Background (CMB), Baryon Acoustic Oscillations (BAO), and Type Ia Supernovae (SNIa). In this section, we summarize the datasets and combinations used in our analysis.
\par
The CMB observations can be further subdivided into two groups: the primary measurements of the temperature and polarization power spectra (binned measurements of the $TT$, $TE$, and $EE$ spectra) and the reconstructed CMB lensing spectrum. The primary CMB measurements are as follows:
\begin{itemize}
    \item Low-$\ell$ ($2\leq\ell<30$) $TT$ power-spectrum data from the \textit{Planck} 2018 \citep{Planck:2019nip} \texttt{Commander} likelihood.
    
    \item Foreground-marginalized, high-$\ell$ ($\ell\geq30$) temperature and $E$-mode polarization data from \textit{Planck} 2018 \citep{Planck:2019nip}, truncated at $\ell<1000$ for $TT$ and $\ell<600$ for $TE$ and $EE$ to combine with the \textit{ACT} DR6 data, following the prescription of~\cite{AtacamaCosmologyTelescope:2025blo}.
    
    \item Foreground-marginalized $T$- and $E$-mode data from the Atacama Cosmology Telescope (\textit{ACT}) DR6 spanning $600<\ell<6500$ \citep{AtacamaCosmologyTelescope:2025blo}.
    
    \item Foreground-marginalized $T$- and $E$-mode data from the South Pole Telescope (\textit{SPT-3G}) D1 \citep{SPT-3G:2025bzu} using the differentiable CMB likelihood framework \texttt{candl} developed by~\citealp{Balkenhol:2024sbv}. The $TT$ spectrum is measured over $400\leq\ell\leq3000$, while the $TE$ and $EE$ spectra cover the range $400\leq\ell\leq4000$.
\end{itemize}

The CMB lensing spectrum datasets consist of quadratic-estimator reconstructions by~\citealp{Carron:2022eyg} on \textit{Planck} PR4 NPIPE maps and on \textit{ACT} DR6 maps by~\citealp{ACT:2023dou} and~\citealp{ACT:2023kun}. We also combine these with the lensing spectrum obtained by~\citealp{SPT-3G:2024atg} from performing the marginal unbiased score expansion (MUSE) algorithm for a map-level Bayesian inference on polarization maps of \textit{SPT-3G} D1.
\par
In addition to the above combinations of CMB datasets, the parameter $\tau_\mathrm{reio}$, which is the optical depth to reionization, is almost entirely constrained by low-$\ell$ $E$-mode data. For this reason, we choose to impose a Gaussian prior on $\tau_\mathrm{reio}\sim\mathcal{N}(0.051,0.006)$, obtained from the analysis of \textit{Planck} NPIPE maps by~\citealp{Planck:2020olo}.
\par
The combination of these datasets is labelled as ``CMB-SPA'' in our analysis and represents the strongest constraints on cosmological models from CMB dataset combinations. The choice of combining the datasets from \textit{Planck}, \textit{ACT}, and \textit{SPT-3G} is justified in~\citealp{SPT-3G:2025bzu}, where it is assumed that, due to the small sky coverage of \textit{SPT-3G}, it is uncorrelated with \textit{ACT} DR6. The correlations between \textit{ACT} and \textit{Planck} are also minimized by using the truncated \textit{Planck} spectra, as stated in~\citealp{AtacamaCosmologyTelescope:2025blo}.
\par
The BAO measurements come from the large-scale structure survey conducted by the Dark Energy Spectroscopic Instrument (\textit{DESI}) DR2 \citep{DESI:2025zgx}, where the BAO signal was measured across the redshift range $0.1<z<4.2$ using various matter tracers. These data are labelled as ``DESI'' in our analysis.
\par
The SNIa measurements are either the \textit{Pantheon+} catalogue by~\citealp{Brout:2022vxf}, labelled as ``PP'' in our work, or the Dark Energy Survey (\textit{DES}) Y5 catalogue \citep{DES:2024jxu}, reanalysed with the updated ``Dovekie'' calibration \citep{Popovic:2025glk} by~\citealp{DES:2025sig}, which we label as ``DD''.

\subsection{Parameter estimation and model comparison}\label{subsec:param estimation}

\begin{table}[htbp]
\centering
\label{tab:priors}
\renewcommand{\arraystretch}{1.3}
\begin{tabular}{lll}
\hline\hline
Parameter & Description & Prior \\
\hline
\multicolumn{3}{l}{\textit{CMB Nuisance Parameters}} \\
\hline
$A_\mathrm{Planck}$ & Planck absolute calibration & $\mathcal{U}(0.5,1.5)$ \\
$A_\mathrm{ACT}$ & ACT absolute calibration & matched to $A_\mathrm{{Planck}}$ \\
$P_\mathrm{ACT}$ & ACT polarization efficiency & $\mathcal{U}(0.9, 1.1)$ \\
$T_\mathrm{cal}$ & SPT-3G D1 global temperature calibration & $\mathcal{N}(1.0, 0.0036)$ \\
$E_\mathrm{cal}$ & SPT-3G D1 global polarization calibration & $\mathcal{U}(0.8, 1.2)$ \\
\hline
\multicolumn{3}{l}{\textit{$\Lambda\mathrm{CDM}$ Parameters}} \\
\hline
$H_0$ & Expansion rate today [km/s/Mpc] & $\mathcal{U}(40,100)$ \\
$100\theta_s$ & Angular size of the sound horizon at recombination & $\mathcal{U}(0.5,10)$ \\
$\Omega_\mathrm{b} h^2$  & Physical baryon density & $\mathcal{U}(0.005, 0.1)$ \\
$\Omega_\mathrm{c} h^2$  & Physical cold dark matter density & $\mathcal{U}(0.001, 0.99)$ \\
$\ln(10^{10}A_\mathrm{s})$ & Amplitude of the scalar primordial power spectrum & $\mathcal{U}(1.61,3.91)$ \\
$n_s$ & Spectral index of the scalar primordial power spectrum & $\mathcal{U}(0.8,1.2)$ \\
$\tau_\mathrm{reio}$ & Optical depth at reionization & $\mathcal{N}(0.051, 0.006)$ \\
\hline
\multicolumn{3}{l}{\textit{$w_0w_a\mathrm{CDM}$ Parameters}} \\
\hline
$w_0$ & DE equation of state parameter for Eq. \ref{eq:CPL} & $\mathcal{U}(-3.0, 1.0)$ \\
$w_a$ & DE equation of state parameter for Eq. \ref{eq:CPL} & $\mathcal{U}(-3.0, 2.0)$ \\
\hline
\multicolumn{3}{l}{\textit{QCD-DE Model Parameters}} \\
\hline
$z_q$ & Location of the $\beta(z)$ switch function & $\mathcal{U}(0.0, 3.0)$ \\
$\Delta z_q$ & Width of the $\beta(z)$ switch function & $\mathcal{U}(0.0, 3.0)$ \\
\hline\hline
\end{tabular}
\caption{Prior distributions for MCMC sampling. The CMB nuisance parameters are sampled only when CMB-SPA is part of the dataset combination. We choose to sample over $\theta_s$ when CMB-SPA is included and over $H_0$ otherwise. If CMB-SPA is not included, $\tau_\mathrm{reio}=0.06$ is fixed.}
\end{table}

We perform Markov chain Monte Carlo (MCMC) sampling using the Metropolis-Hastings algorithm (implemented by~\citealp{Lewis:2002ah,Lewis:2013hha}) via the publicly available Code for Bayesian Analysis (\texttt{Cobaya}) developed by~\citealp{Torrado:2020dgo}. We deem the chains to be converged when the Gelman-Rubin statistic \citep{Gelman:1992zz} satisfies $R-1<0.01$.
\par
We perform MCMC sampling for the QCD-DE models, alongside the baseline $\Lambda\mathrm{CDM}$ and $w_0w_a\mathrm{CDM}$ models for comparison, using the following dataset combinations: PP+DESI, CMB-SPA, and CMB-SPA+PP+DESI. In addition, we choose to sample the DD+DESI and CMB-SPA+DD+DESI combinations to include the SN sample with the latest calibration methods, as described by~\citealp{Popovic:2025glk}.
\par
The priors on the CMB nuisance parameters, as well as the cosmological model parameters are listed on Tab.~(\ref{tab:priors}).
\par
For a given set of datasets, we can perform model comparisons using a variety of methods. The most popular method of comparison in the cosmology community, given the MCMC chains of a set of models, is to obtain the difference in the minimum $\chi^2_\mathrm{eff}\equiv-2\ln\mathcal{L}$, where $\mathcal{L}\equiv P(d\vert\theta,\mathcal{M})$ is the likelihood of the observed data $d$ given a set of parameters $\theta$ for a specific model $\mathcal{M}$. We find the maximum of the posterior distribution (MAP) using the Bound Optimization by Quadratic Approximation (\texttt{BOBYQA}) algorithm as implemented by~\citealp{Cartis:2018xum,Cartis:2018jxl}. The difference in $\chi^2_\mathrm{MAP}$ between the alternative model and the baseline $\Lambda\mathrm{CDM}$ model provides an indication of how well the best-fit model describes the data for a given dataset combination. For $\Delta\chi^2 = \chi^2_\mathrm{alt}-\chi^2_{\Lambda\mathrm{CDM}}$, if $\Delta\chi^2<0$ it indicates that the alternative model is preferred, while $\Delta\chi^2>0$ indicates a preference for $\Lambda\mathrm{CDM}$.
\par
For models such as $w_0w_a\mathrm{CDM}$, which contain the baseline $\Lambda\mathrm{CDM}$ within their parameter space ($w_0=-1$, $w_a=0$), $\Delta\chi^2\leq0$ by construction. For such cases of nested models, we can use Wilks' theorem \citep{Wilks:1938dza}, which states that for nested models, $\Delta\chi^2$ follows a $\chi^2$ distribution with degrees of freedom equal to the number of additional parameters in the larger model (for $w_0w_a\mathrm{CDM}$ this is equal to 2). However, by the definition of our QCD-DE model, it behaves like $\Lambda\mathrm{CDM}$ at early times, but always deviates from $\Lambda\mathrm{CDM}$ at late times. Therefore, we cannot directly quantify preferences for or against the QCD-DE model in terms of standard deviations using Wilks' theorem.
\par
A possible resolution to this is to compute the Akaike Information Criterion (AIC), defined as
\begin{equation}\label{AIC}
    \mathrm{AIC} = \chi^2_\mathrm{ML}+2k
\end{equation}
where $\chi^2_\mathrm{ML}$ is the effective $\chi^2$ evaluated at the maximum likelihood and $k$ is the number of model parameters \citep{akaikeNewLookStatistical1974}. For the difference
\begin{equation}
    \Delta\mathrm{AIC} = \mathrm{AIC}_\mathrm{alt}-\mathrm{AIC}_{\Lambda\mathrm{CDM}},
\end{equation}
a negative value indicates that the alternative model is preferred.
\par
Another information criterion, introduced by~\citealp{Spiegelhalter:2002yvw}, is the Deviance Information Criterion (DIC). As described in~\citealp{gelmanBayesianDataAnalysis2013}, it can be thought of as a ``somewhat Bayesian'' version of the AIC. It is defined as
\begin{equation}\label{DIC}
    \mathrm{DIC} = \chi^2(\bar{\theta})+2p_D = \overline{\chi^2(\theta)}+p_D
\end{equation}
where $\bar{\theta}$ and $\overline{\chi^2(\theta)}$ represent the posterior means of the parameters and $\chi^2$, respectively, and $p_D$ is the effective number of parameters of the model (Bayesian complexity), given by
\begin{align*}
    p_D &= \overline{\chi^2(\theta)}-\chi^2(\bar{\theta})\\
    &= \frac{1}{2}\mathrm{Var}\left[\chi^2(\theta)\right].
\end{align*}
The second line is an alternative definition of $p_D$ given by~\citealp{gelmanBayesianDataAnalysis2013}, which requires the variance of $\chi^2$ over the posterior distribution. The model comparison based on $\Delta\mathrm{DIC} = \mathrm{DIC}_\mathrm{alt}-\mathrm{DIC}_{\Lambda\mathrm{CDM}}$ follows the same logic as the AIC, such that the preferred model is the one that minimizes the DIC.
\par
As mentioned in~\citealp{Liddle:2007fy}, while the AIC and DIC are motivated by information theory and therefore have a strong statistical motivation, they are also dimensionally inconsistent. This is to say that these information criteria may exhibit a preference for models with more parameters, as the penalty for model overcomplexity is not severe enough to ensure that, in the limit of arbitrarily large datasets, the true model is always preferred.
\par
For a fully Bayesian model comparison, we must compute the Bayesian evidence, given by
\begin{equation}
    \mathcal{Z} = P(d\vert\mathcal{M}) = \int\mathrm{d}\theta\,\mathcal{L}(\theta)\pi(\theta),
\end{equation}
where $\pi(\theta)$ is the prior distribution. If we assume equal prior probabilities for the models, the relative posterior probabilities of two models are given by the ratio of their Bayesian evidences. We can therefore define the (log) Bayes factor as
\begin{equation}\label{Bayes factor}
    \ln\mathcal{B} = \ln\mathcal{Z}_\mathrm{alt}-\ln\mathcal{Z}_{\Lambda\mathrm{CDM}}
\end{equation}
such that $\ln\mathcal{B}>0$ indicates a preference for the alternative model, while $\ln\mathcal{B}<0$ indicates a preference for $\Lambda\mathrm{CDM}$. The problem for cosmological models is that the computation of the Bayesian evidence is highly non-trivial and, for an accurate computation, approaches such as nested sampling are required to probe the full extent of the prior distribution. However, attempts have been made to compute the Bayesian evidence from MCMC chains, such as the nearest-neighbours algorithm implemented by~\citealp{Heavens:2017afc} in the code \texttt{MCEvidence}.
\par
In this paper, we use the method developed by~\citealp{mcewenMachineLearningAssisted2023}, which uses the harmonic mean estimator of $\mathcal{Z}^{-1}$ written as
\begin{equation}\label{lhme}
    \widehat{\mathcal{Z}^{-1}} = \frac{1}{N}\sum^N_{i=1}\frac{\varphi(\theta_i)}{\mathcal{L}(\theta_i)\pi(\theta_i)}, ~~~ \theta_i\sim p(\theta\vert d)
\end{equation}
where $\varphi(\theta)$ is an arbitrary normalized distribution, which is required to have thinner tails than the posterior for the estimator to provide a reliable approximation of the evidence.
\par
The difficulty in choosing $\varphi$ is resolved by~\citealp{Polanska:2024arc}, who use normalizing flows, in which a machine-learning model is trained on a subset of posterior samples to produce a normalized distribution that approximates the posterior. The learned distribution is concentrated by lowering its ``temperature'' such that it is contained within the posterior. For the learned harmonic mean estimator (LHME) and normalizing flows, we used the publicly available code \texttt{harmonic}\footnote{\url{https://github.com/astro-informatics/harmonic}} developed by~\citealp{mcewenMachineLearningAssisted2023}. 
\par
For the analysis presented in this paper, we developed a \texttt{Python} package, \texttt{CosmoCAT},\footnote{\url{https://github.com/dlehdgk/CosmoCAT}} which contains a wrapper for \texttt{harmonic} to be used with cosmological MCMC chains generated by \texttt{Cobaya}, alongside all the other comparison metrics described above.

\section{Results}
\label{Sec:resuts}

\subsection{Parameter Constraints and Statistical Evidence}\label{subsec:statistical_evidence}

The constraints on $\{H_0, \Omega_m, \sigma_8, S_8\}$ are shown in Tab.~(\ref{tab:results}), together with the QCD-DE parameters $\{z_q, \Delta z_q\}$ and the asymptotic Hubble parameter $\overline{H}$. For $\Lambda\mathrm{CDM}$, we define $\overline{H}\equiv\sqrt{\frac{8\pi G}{3}\rho_\Lambda}$ to facilitate a direct comparison with the QCD-DE models.
The full list of cosmological parameter constraints for the QCD-DE model can be found in Appendix (\ref{ap:constraints})
\par
\begin{figure}
    \centering
    \includegraphics[width=0.8\linewidth]{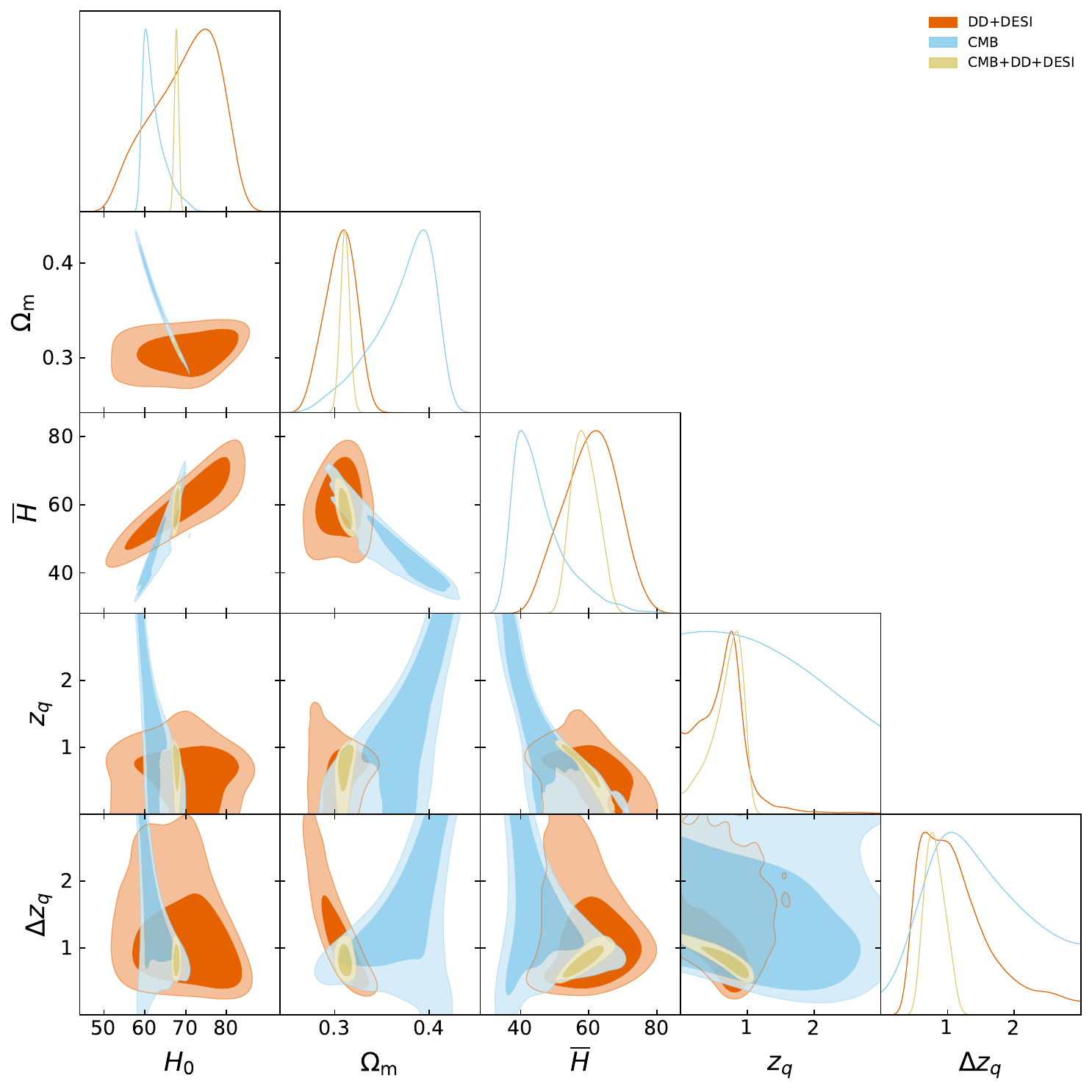}
    \caption{Triangle plot of the constraints on the QCD-DE model given different dataset combinations.}
    \label{fig:datasets}
\end{figure}

We observe from Fig.~(\ref{fig:datasets}) that the CMB data favour larger values of $\Omega_m = 0.373^{+0.040}_{-0.016}$ compared to the late-time data, which prefer $\Omega_m = 0.306^{+0.019}_{-0.015}$ in the QCD-DE model. However, due to the degeneracy between $\Omega_m$ and the QCD-DE parameters, the posterior distribution exhibits a pronounced low-$\Omega_m$ tail, extending to $\Omega_m = 0.373^{+0.050}_{-0.070}$ at the $95\%$ CL.
\par
We also find that the CMB data alone provide only an upper bound on the transition redshift, $z_q<1.77$ at the $68\%$ CL. The addition of late-time distance measurements significantly improves the constraints, yielding $z_q = 0.68^{+0.31}_{-0.14}$.
\par
However the same cannot be said for $\Delta z_q$, where the CMB alone is able to impose constraints on $\Delta z_q = 1.50^{+0.66}_{-0.95}$. The late-time distance measurements appear to prefer lower values of $\Delta z_q = 0.64^{+0.30}_{-0.37}$, and combining the CMB with distance measurements is able to impose a strong constraint on $\Delta z_q=0.82^{+0.13}_{-0.18}$.
\par
\begin{figure}
    \centering
    \includegraphics[width=0.65\linewidth]{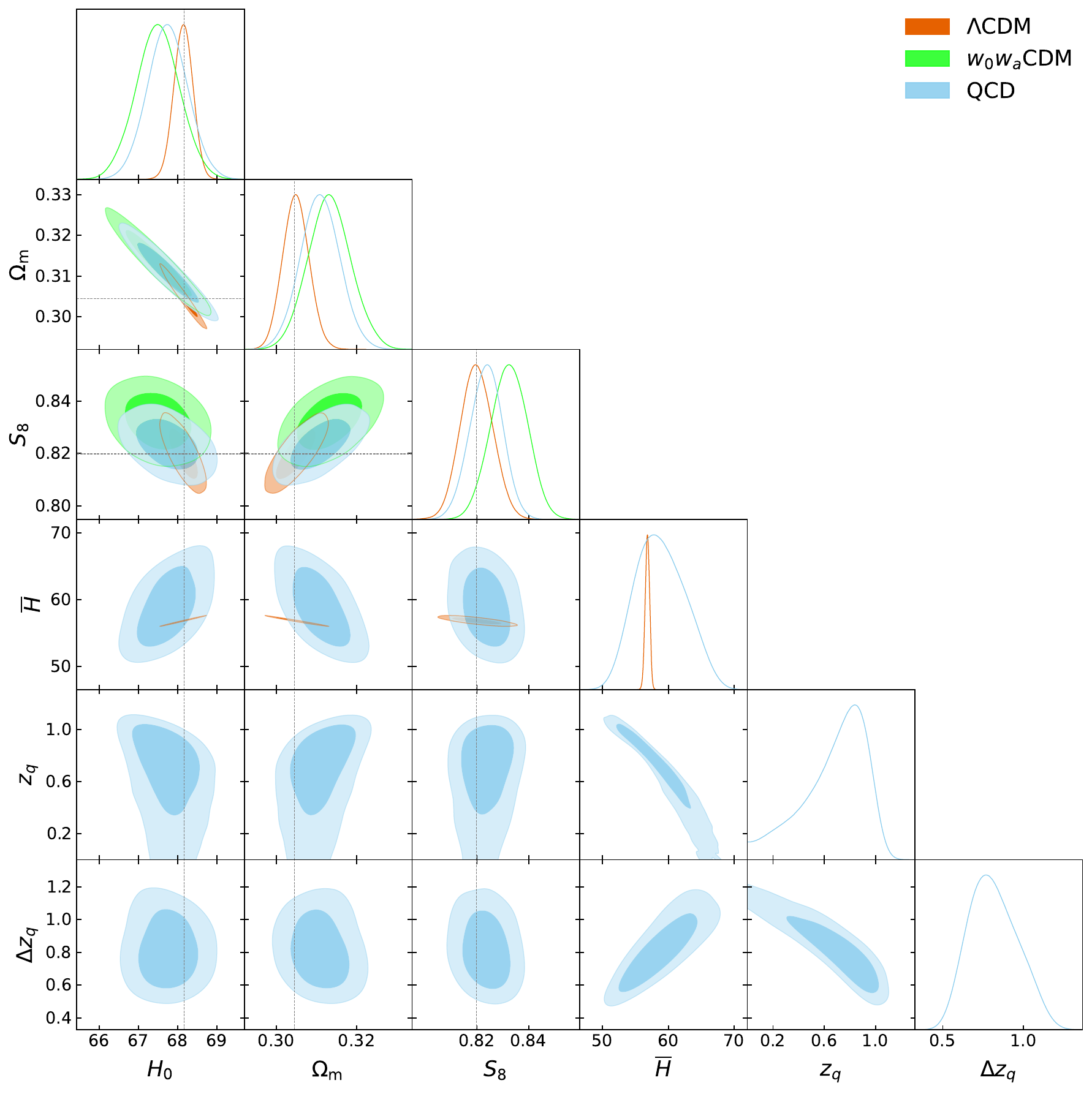}
    \caption{Triangle plot of cosmological parameter constraints for $\Lambda\mathrm{CDM}$, $w_0w_a\mathrm{CDM}$, and the QCD-DE model obtained from the CMB-SPA+DD+DESI dataset combination. The corresponding constraints obtained with the CMB-SPA+PP+DESI dataset combination are fully consistent with those derived using DD for all models considered and thus not shown.}
    \label{fig:triangle plot}
\end{figure}

As shown in Fig.~(\ref{fig:triangle plot}), the introduction of additional degrees of freedom associated with dynamical dark energy broadens the allowed parameter space, particularly in the $H_0-\Omega_m$ plane, similarly to what is observed in the $w_0w_a\mathrm{CDM}$ model.
When combining all the datasets, the QCD-DE model shifts the inferred value of $H_0 = 67.74\pm 0.52$ towards lower values relative to $\Lambda\mathrm{CDM}$ ($68.14\pm 0.24$). This behaviour is analogous to that found in conventional dynamical dark energy models and can be understood as a consequence of the geometrical degeneracy. The additional freedom in the late-time expansion history allows the model to preserve the distance to last scattering preferred by the CMB while accommodating different values of the present-day Hubble constant.
\par
This shift in $H_0$ is accompanied by a corresponding increase in $\Omega_m$. The origin of this behaviour is twofold. First, the CMB tightly constrains the physical matter density $\Omega_m h^2$, leading to the well-known geometrical degeneracy between $H_0$ and $\Omega_m$. Second, the supernova datasets generally favour slightly larger values of $\Omega_m$, allowing the model to partially compensate for the modified expansion history introduced by the QCD-DE component. As a result, the QCD-DE model tends to occupy an intermediate position between $\Lambda\mathrm{CDM}$ and $w_0w_a\mathrm{CDM}$ in the parameter space.
\par
Finally, we find that replacing the \textit{Pantheon+} sample with the recalibrated \textit{DES}-Dovekie sample has a negligible impact on the inferred cosmological parameters. The constraints obtained using PP and DD are fully consistent with each other for all models considered, indicating that the conclusions of this work are robust with respect to the choice of supernova compilation.
\begin{table*}
    \renewcommand{\arraystretch}{1.3}
    \centering
    \label{tab:results}
    \makebox[\linewidth][c]{%
    \resizebox{0.95\textwidth}{!}{%
    \begin{tabular}{c | ccc | ccc}
        \hline\hline
        \multirow{2}{*}{Parameter}
        & \multicolumn{3}{c|}{CMB-SPA+PP+DESI}
        & \multicolumn{3}{c}{CMB-SPA+DD+DESI} \\
        \cline{2-7}
        & $\Lambda\mathrm{CDM}$ & $w_0w_a\mathrm{CDM}$ & QCD-DE
        & $\Lambda\mathrm{CDM}$ & $w_0w_a\mathrm{CDM}$ & QCD-DE \\
        \hline
        $H_0$ & $68.15\pm 0.25$ & $67.68\pm 0.60$ & $67.74\pm 0.58$ & $68.14\pm 0.24$ & $67.49\pm 0.55$ & $67.74\pm 0.52$ \\
        $\Omega_m$ & $0.3046\pm 0.0033$ & $0.3113\pm 0.0057$ & $0.3109\pm 0.0054$ & $0.3049\pm 0.0033$ & $0.3132\pm 0.0053$ & $0.3109\pm 0.0049$ \\
        $\sigma_8$ & $0.8131\pm 0.0038$ & $0.8156\pm 0.0074$ & $0.8091\pm 0.0060$ & $0.8133\pm 0.0039$ & $0.8147\pm 0.0071$ & $0.8090\pm 0.0056$ \\
        $S_8$ & $0.8193\pm 0.0063$ & $0.8308\pm 0.0071$ & $0.8235\pm 0.0064$ & $0.8198\pm 0.0063$ & $0.8324\pm 0.0071$ & $0.8235\pm 0.0063$ \\
        \hline
        $z_q$ & $---$ & $---$ & $0.66^{+0.32}_{-0.16}$ & $---$ & $---$ & $0.68^{+0.31}_{-0.14}$ \\
        $\Delta z_q$ & $---$ & $---$ & $0.83^{+0.14}_{-0.18}$ & $---$ & $---$ & $0.82^{+0.13}_{-0.18}$ \\
        $\overline{H}$ & $56.83\pm 0.33$ & $---$ & $59.1^{+3.8}_{-4.4}$ & $56.81\pm 0.33$ & $---$ & $58.8^{+3.5}_{-4.3}$ \\
        \hline\hline
    \end{tabular}
    }
    }
    \caption{Constraints on the cosmological parameters of $\Lambda\mathrm{CDM}$, $w_0w_a\mathrm{CDM}$ and the QCD-DE models from combinations of CMB-SPA, DESI DR2 BAO, and SNIa data from either the DD or PP surveys. 
    }
\end{table*}
\par
\begin{figure}
    \centering
    \includegraphics[width=0.8\linewidth]{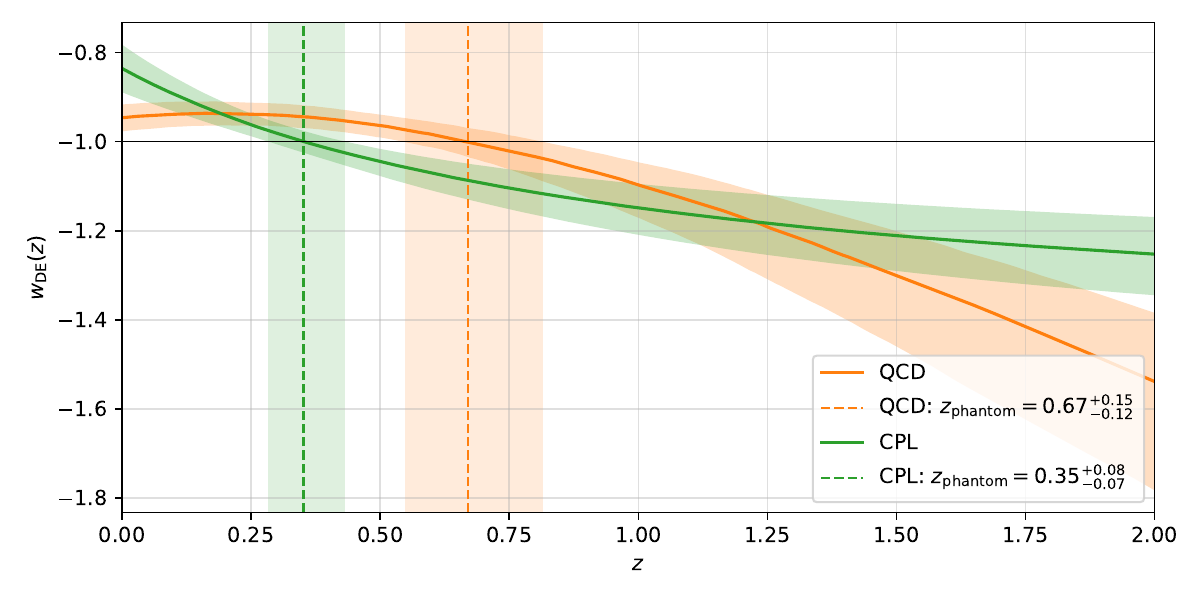}
    \caption{
    Posterior reconstruction of the effective QCD-DE equation of state and the CPL equation of state obtained from the constraints of the CMB-SPA+DD+DESI dataset combination.
    }
    \label{fig:eos posterior}
\end{figure}

With constraints from observations, we can attempt to reconstruct the posterior of the DE EoS of the CPL and QCD-DE models. Fig.~(\ref{fig:eos posterior}) shows the functional posterior of the effective EoS of the QCD-DE model alongside the EoS posterior given by the CPL parametrization. We note that the QCD-DE models prefer an earlier effective phantom crossing of DE when compared with the CPL parametrization~\citep{Ozulker:2025ehg} and a present-day value of the effective equation of state closer to $-1$.
\par
This is a particularly interesting result. A large number of dynamical dark energy parametrizations have been explored in the literature and, despite the different functional forms adopted, the preferred phantom-crossing redshift has generally been found to be remarkably stable around $z\sim0.3-0.4$~\citep{Giare:2024gpk}. In contrast, the QCD-DE model favours a substantially earlier crossing, around $z\sim0.67$. This behaviour appears to be a distinctive feature of the QCD-DE framework and may reflect the different physical origin of the effective dark energy evolution. Rather than arising from the dynamics of a propagating degree of freedom such as a scalar field, the evolution is driven by the time dependence of the vacuum-energy correction associated with the QCD-induced mechanism. Despite this significantly earlier phantom crossing, the model remains consistent with the latest reconstruction of the DE EoS by \citealp{Kessler:2026dbi} which appears to also prefer an earlier phantom crossing at $z\sim0.7$.
\par
While the description of the evolution is specific to this particular parametrization, we expect to observe similar behaviour in other forms which attempt to faithfully describe the effects of the QCD vacuum. However, we leave other parametrization and extensions to further work.
\par
\begin{figure}
    \centering
    \includegraphics[width=0.45\linewidth]{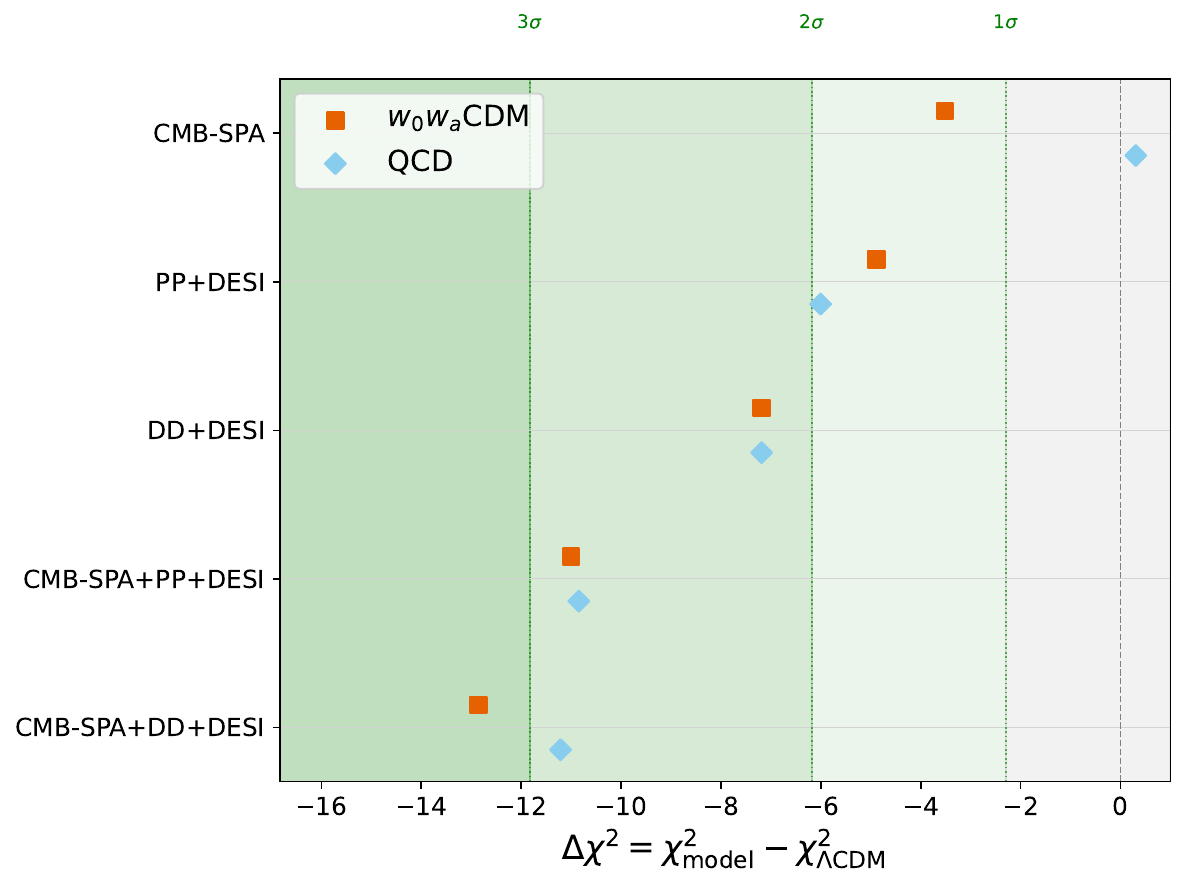}
    \includegraphics[width=0.45\linewidth]{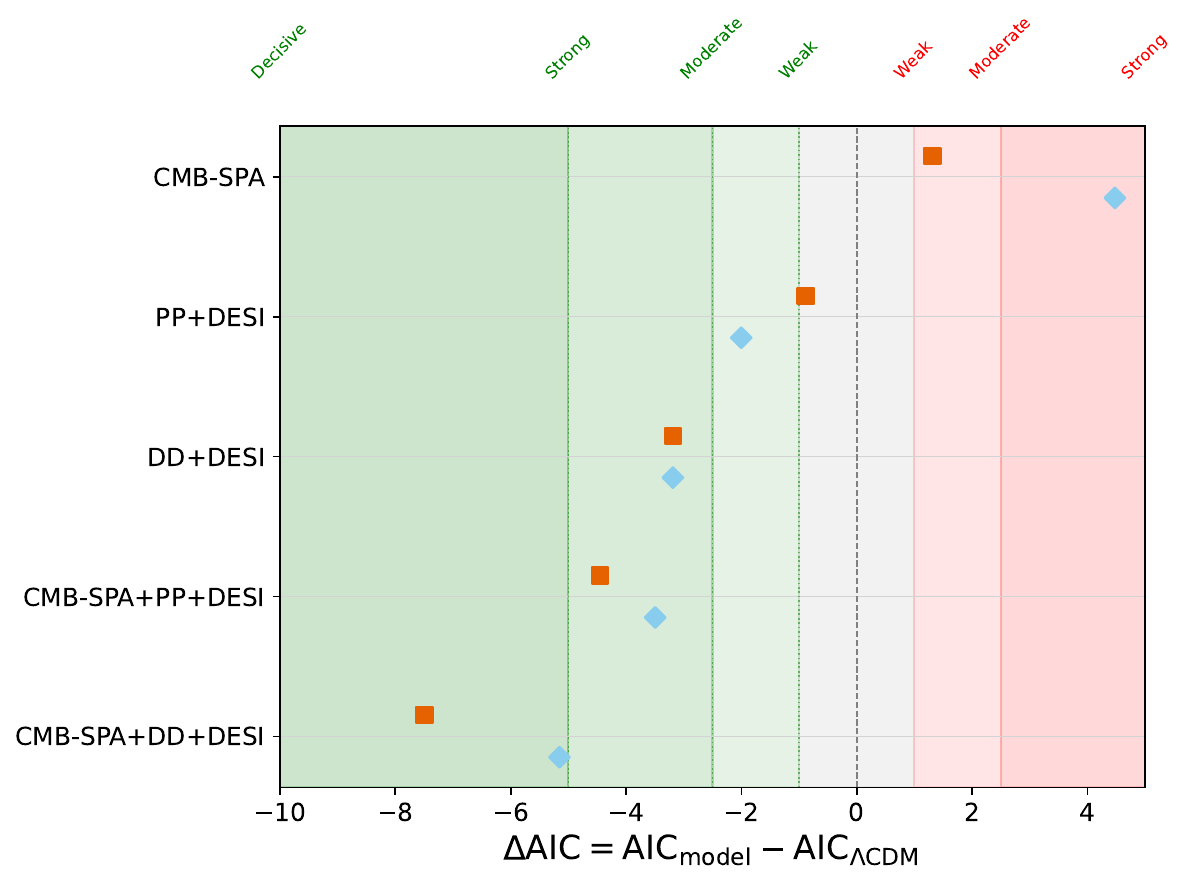}
    \includegraphics[width=0.45\linewidth]{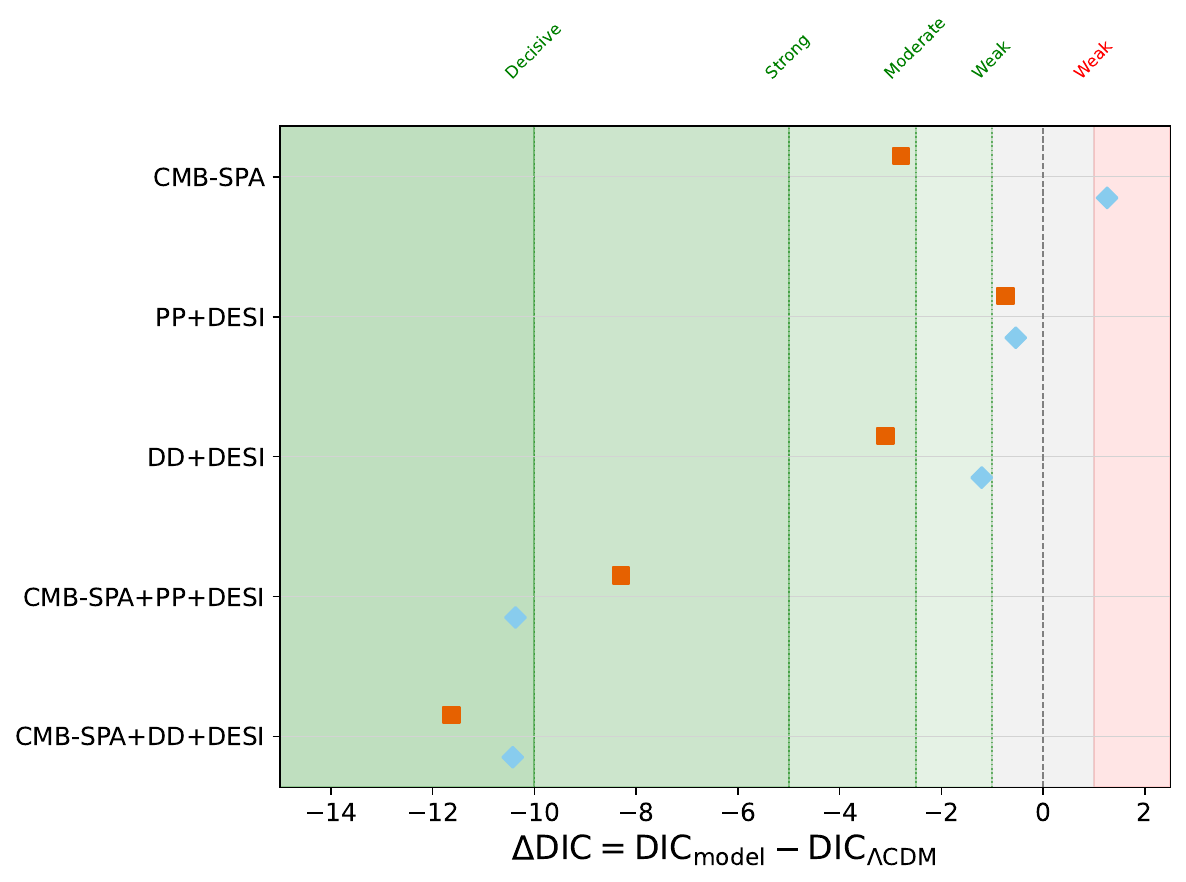}
    \includegraphics[width=0.45\linewidth]{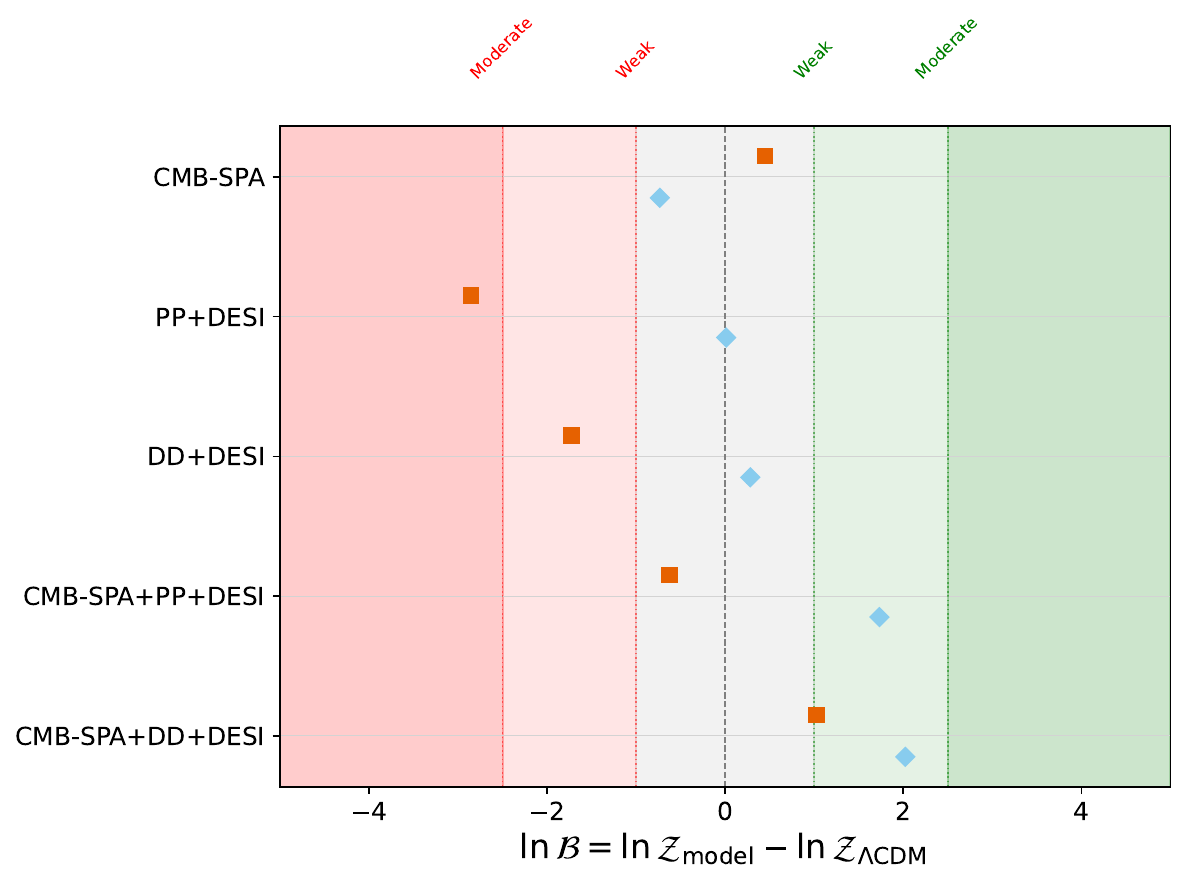}
    \caption{(Top left) Plot of the difference in the MAP $\chi^2_\mathrm{eff}$ with respect to the $\Lambda\mathrm{CDM}$ model. The contours indicate the standard deviations for a $\chi^2$ distribution with two degrees of freedom, corresponding to the expected distribution of $\Delta\chi^2=\chi^2_{w_0w_a}-\chi^2_\Lambda$ according to Wilks' theorem \citep{Wilks:1938dza}. The contours are not applicable to QCD-DE since it is not a direct extension of $\Lambda\mathrm{CDM}$.
    (Top right) Plot of the $\Delta\mathrm{AIC}$ with respect to the $\Lambda\mathrm{CDM}$ model.
    (Bottom left) Plot of $\Delta\mathrm{DIC}$ using $p_D = \frac{1}{2}\mathrm{Var}(\chi^2_\mathrm{eff})$ to evaluate the DIC.
    (Bottom right) Plot of the Bayes factors with respect to the $\Lambda\mathrm{CDM}$ evidence. The contours for $\Delta\mathrm{AIC}$, $\Delta\mathrm{DIC}$ and $\ln\mathcal{B}$ indicate the boundaries of the Jeffreys scale as described by \citep{Trotta:2008qt}. For $\Delta\chi^2$, $\Delta\mathrm{AIC}$ and $\Delta\mathrm{DIC}$, a negative value indicates a preference for the alternative model over $\Lambda\mathrm{CDM}$ while for $\ln\mathcal{B}$, it is the positive. The values are listed in the Appendix (\ref{ap:Model comparison}).}
    \label{fig:chi2}
\end{figure}
As described in Section~\ref{subsec:param estimation}, we use a variety of model-comparison metrics to quantify the performance of the QCD-DE model relative to $\Lambda\mathrm{CDM}$ and $w_0w_a\mathrm{CDM}$. We begin with the goodness-of-fit, where the $\chi^2_\mathrm{eff}$ values obtained for each model and dataset combination are shown in the top-left panel of Fig.~(\ref{fig:chi2}). It is evident that the QCD-DE models perform similarly to $w_0w_a\mathrm{CDM}$ for most dataset combinations, with the largest differences appearing for the CMB-only datasets. As expected, $w_0w_a\mathrm{CDM}$ always performs at least as well as $\Lambda\mathrm{CDM}$ because the latter is recovered in the limit $w_0=-1$ and $w_a=0$.
\par
We note that $\Lambda\mathrm{CDM}$ contains two fewer free parameters than either of the dynamical dark energy models. We therefore also consider the $\Delta\mathrm{AIC}$ and $\Delta\mathrm{DIC}$ statistics, which penalize additional model complexity. These results are shown in the top-right and bottom-left panels of Fig.~(\ref{fig:chi2}), respectively. Although neither criterion is fully Bayesian and neither imposes a sufficiently strong penalty to guarantee selection of the true model in the limit of infinite data, both provide useful complementary measures of model performance.
\par
We also present the Bayes factors estimated using \texttt{harmonic}, shown in the bottom-right panel of Fig.(\ref{fig:chi2}). For $w_0w_a\mathrm{CDM}$, we recover the preference for $\Lambda\mathrm{CDM}$ for all dataset combinations except those involving the combination of CMB data with the DES-Dovekie supernova sample, consistent with the findings of \citep{Ong:2026tta}. The QCD-DE models display a qualitatively different behaviour. The QCD-DE parametrization remains more competitive in the Bayesian evidence for all the dataset combinations considered. For the CMB-only and late-time-only datasets, the QCD-DE is statistically indistinguishable from $\Lambda\mathrm{CDM}$, while for the full dataset combinations they are mildly preferred over the standard model.
\par
A noteworthy feature is the stability of the QCD-DE results with respect to the choice of supernova sample. The $\Delta\chi^2$, $\Delta\mathrm{AIC}$, $\Delta\mathrm{DIC}$ and $\ln\mathcal{B}$ values obtained for $w_0w_a\mathrm{CDM}$ exhibit a larger variation between the DES-Dovekie and Pantheon+ datasets, whereas the parametrization of QCD-DE is able to remain remarkably consistent across the two supernova compilations.
\par
Taken together, the Bayesian evidence provides a particularly encouraging result for the QCD-DE scenario. Unlike $w_0w_a\mathrm{CDM}$, which often suffers a substantial Bayesian penalty for its additional freedom, the QCD-DE remains competitive with $\Lambda\mathrm{CDM}$ for all dataset combinations and are mildly favoured when the full cosmological dataset is considered. Importantly, this conclusion is consistent with the behaviour observed in $\Delta\chi^2$, $\Delta\mathrm{AIC}$ and $\Delta\mathrm{DIC}$, indicating that all model-selection metrics considered in this work point towards the same overall conclusion: the QCD-DE framework provides a statistically viable, and in some cases preferred, description of the current cosmological observations.

\subsection{CMB angular power spectra}\label{subsec:power spectra}

\begin{figure}
    \centering
    \includegraphics[width=0.9\linewidth]{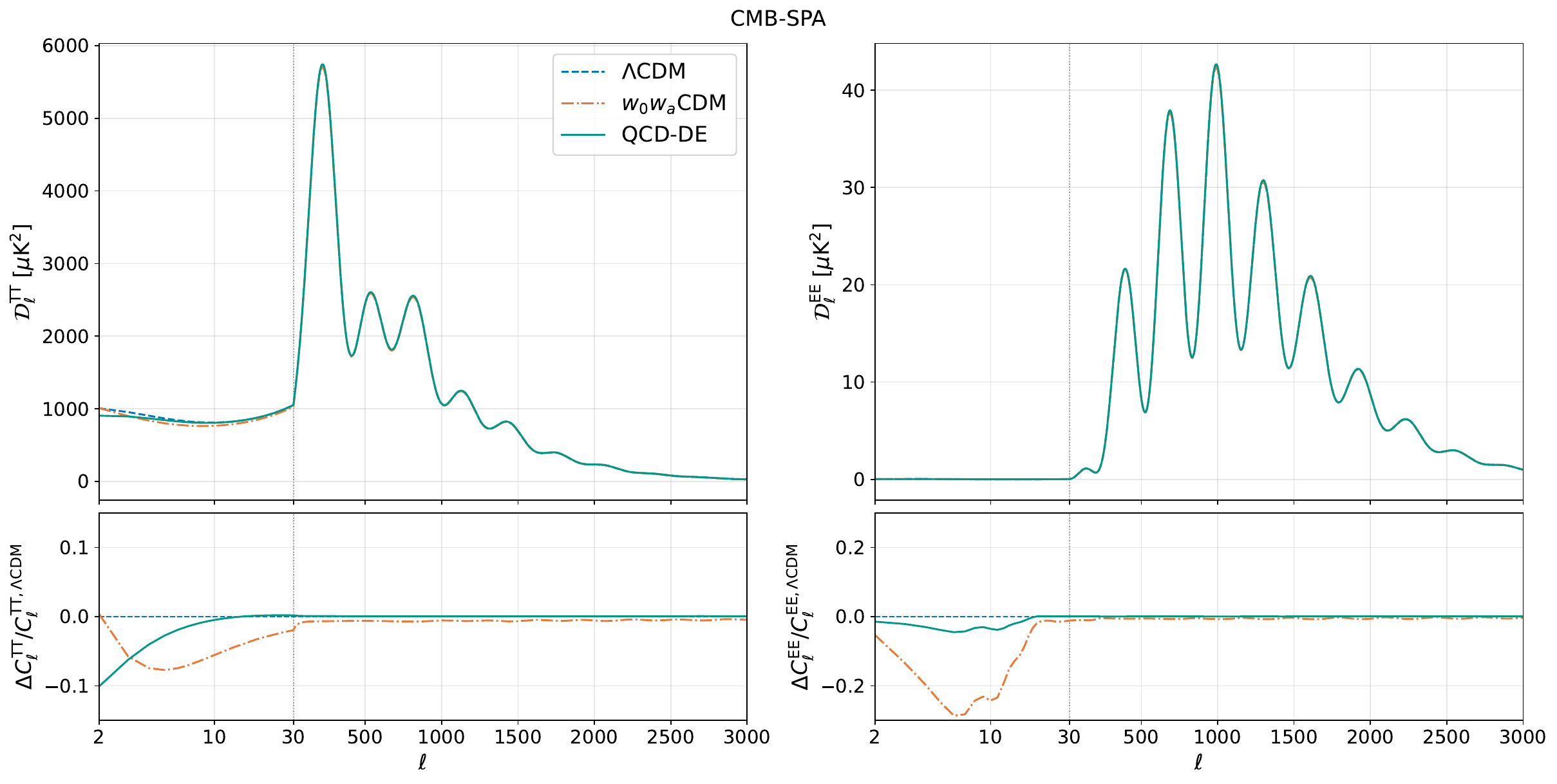}
    \includegraphics[width=0.9\linewidth]{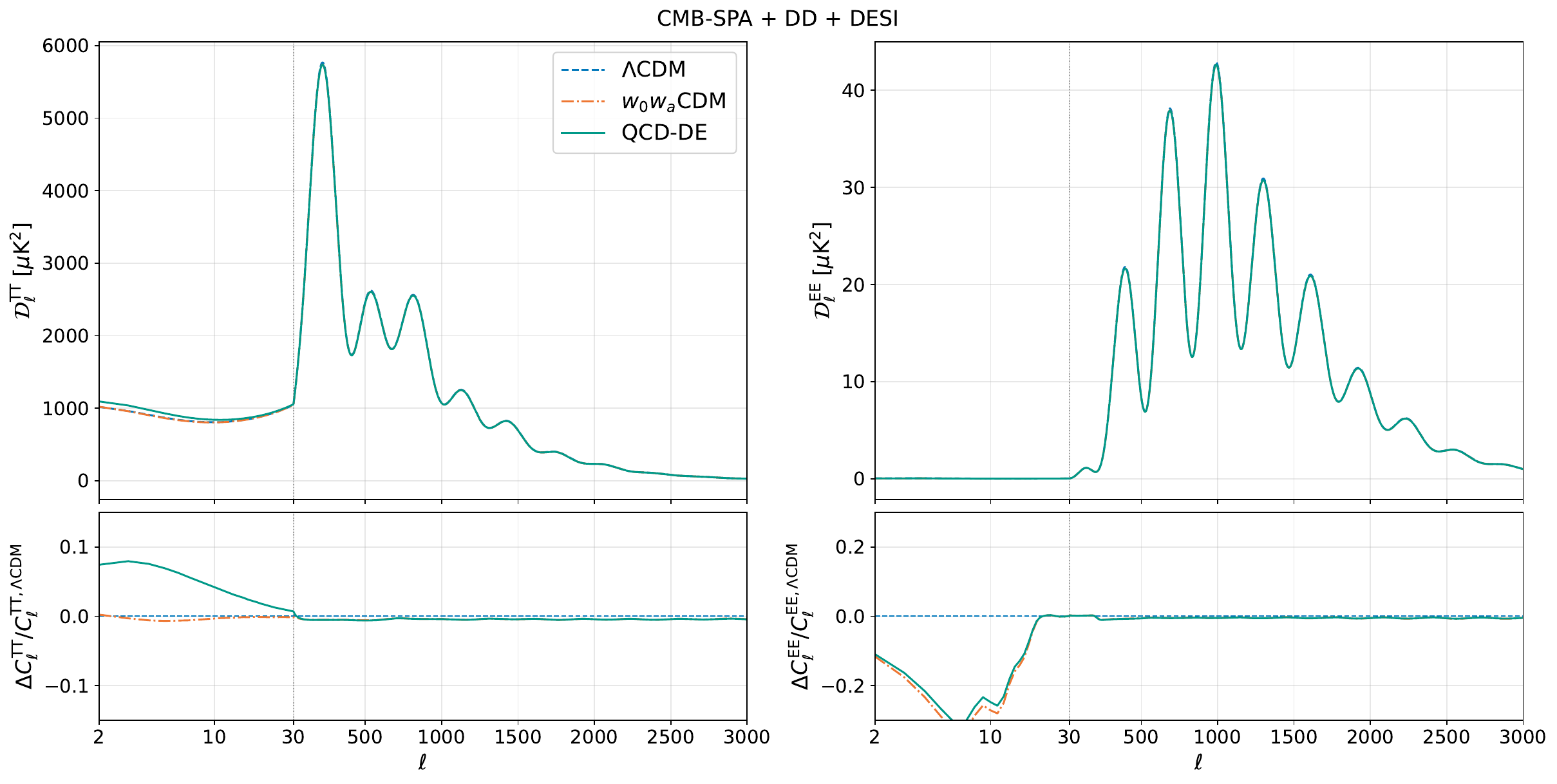}
    \includegraphics[width=0.9\linewidth]{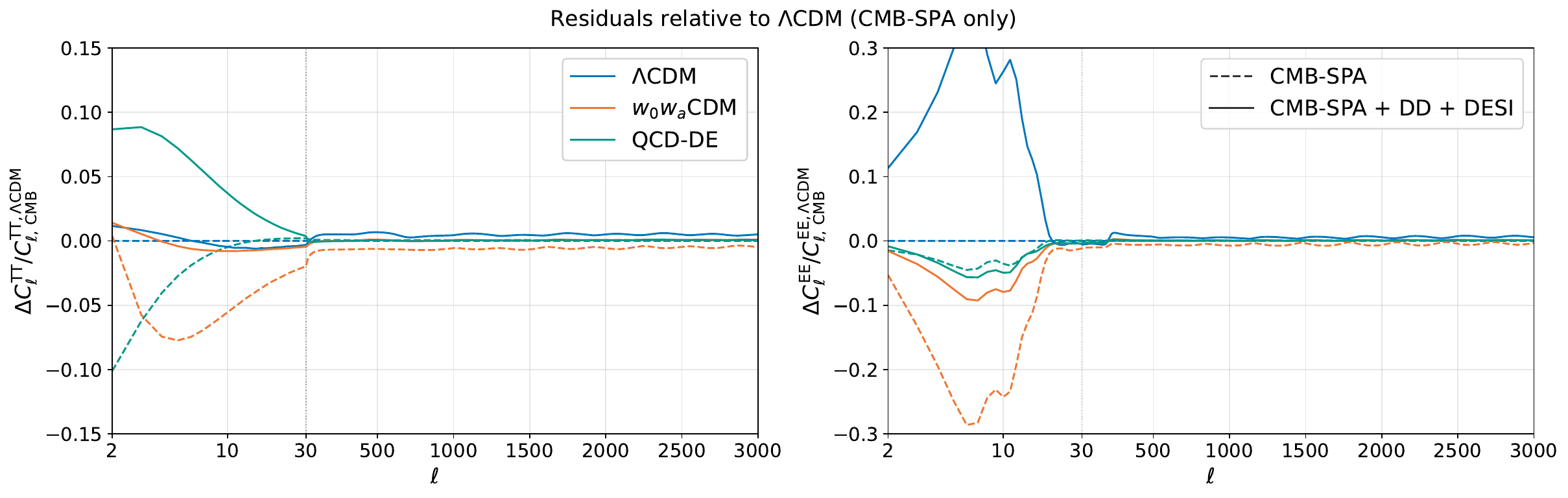}
    \caption{CMB $TT$ and $EE$ angular power spectra and their fractional differences relative to the best-fit $\Lambda\mathrm{CDM}$ model, $\Delta C_\ell/C_\ell^{\Lambda\mathrm{CDM}}$, computed at the MAP cosmological parameters for the CMB-SPA dataset (top panels) and the CMB-SPA+DD+DESI dataset combination (middle panels). (bottom panels) The fractional differences of all the power spectra above, relative to the MAP of $\Lambda\mathrm{CDM}$ for the CMB-SPA dataset.}
    \label{fig:power spectra full}
\end{figure}
To better understand the physical origin of the differences between the models, in Fig.~(\ref{fig:power spectra full}) we compare the CMB $TT$ and $EE$ angular power spectra computed at the MAP cosmological parameters for $\Lambda\mathrm{CDM}$, $w_0w_a\mathrm{CDM}$ and QCD-DE. The figure also shows the fractional differences relative to the corresponding $\Lambda\mathrm{CDM}$ best fit.

For the CMB-SPA dataset alone, all models produce nearly identical predictions over most of the multipole range. In particular, the differences in the damping-tail region are very small, reflecting the fact that the high-$\ell$ anisotropies are strongly constrained by the CMB data and largely insensitive to the details of the late-time dark energy evolution. The largest deviations appear at low multipoles, where late-time effects become important and modifications to the late Integrated Sachs-Wolfe (ISW) contribution can affect the temperature anisotropies.

In the $TT$ spectrum, both the $w_0w_a\rm CDM$ and QCD-DE modify the low-$\ell$ behaviour relative to $\Lambda\mathrm{CDM}$. However, the QCD-DE model provide a larger suppression of power at the lowest multipoles and are able to reproduce the observed low quadrupole more closely than $w_0w_a\rm CDM$. This is one of the main differences between the two dynamical dark energy scenarios and suggests that the modified late-time evolution in the QCD-DE framework has a non-trivial impact on the largest observable scales.

The behaviour in the $EE$ spectrum is also interesting. At low multipoles, while the $w_0w_a\rm CDM$ and QCD-DE models both exhibit a suppression of power relative to $\Lambda\mathrm{CDM}$, the effect is far less pronounced for the QCD-DE best fit in comparison to the CPL. Although the effect remains small in absolute terms, it provides an additional observational signature capable of distinguishing the two classes of models.

When the full dataset combination CMB-SPA+DD+DESI is considered, the situation changes significantly. The $w_0w_a\rm CDM$ model remains qualitatively similar to the $\Lambda\mathrm{CDM}$ case with minor deviations at low multipoles. In contrast, the best-fitting QCD-DE model appears to have an excess of power at low $\ell$ for the $TT$ spectrum in comparison to $\Lambda\mathrm{CDM}$. We also observe a slight suppression of power throughout the damping-tail region, mimicking the behaviour of $w_0w_a\rm CDM$.

For the $EE$ spectrum, the behaviour also changes compared to the CMB-only case. In this case, the QCD-DE and $w_0w_a\rm CDM$ models become qualitatively similar, both showing significant suppression of power relative to $\Lambda\mathrm{CDM}$ at low multipoles. Thus, while the polarization spectra provide a clear distinction between the models in the CMB-only analysis, the inclusion of late-time information drives the preferred QCD-DE solutions towards a behaviour more closely resembling $w_0w_a\rm CDM$ in $EE$, even though significant differences remain in the temperature anisotropies. We can observe from the bottom residual plots in Fig.~(\ref{fig:power spectra full}) that in fact, the preferred fit for the $EE$ spectra across the datasets for QCD-DE remains fairly consistent and it is the $\Lambda\mathrm{CDM}$ reference line which is changing significantly between datasets.

\subsection{BAO residual plots}\label{subsec:BAO residual}

\begin{figure}
    \centering
    \includegraphics[width=0.5\linewidth]{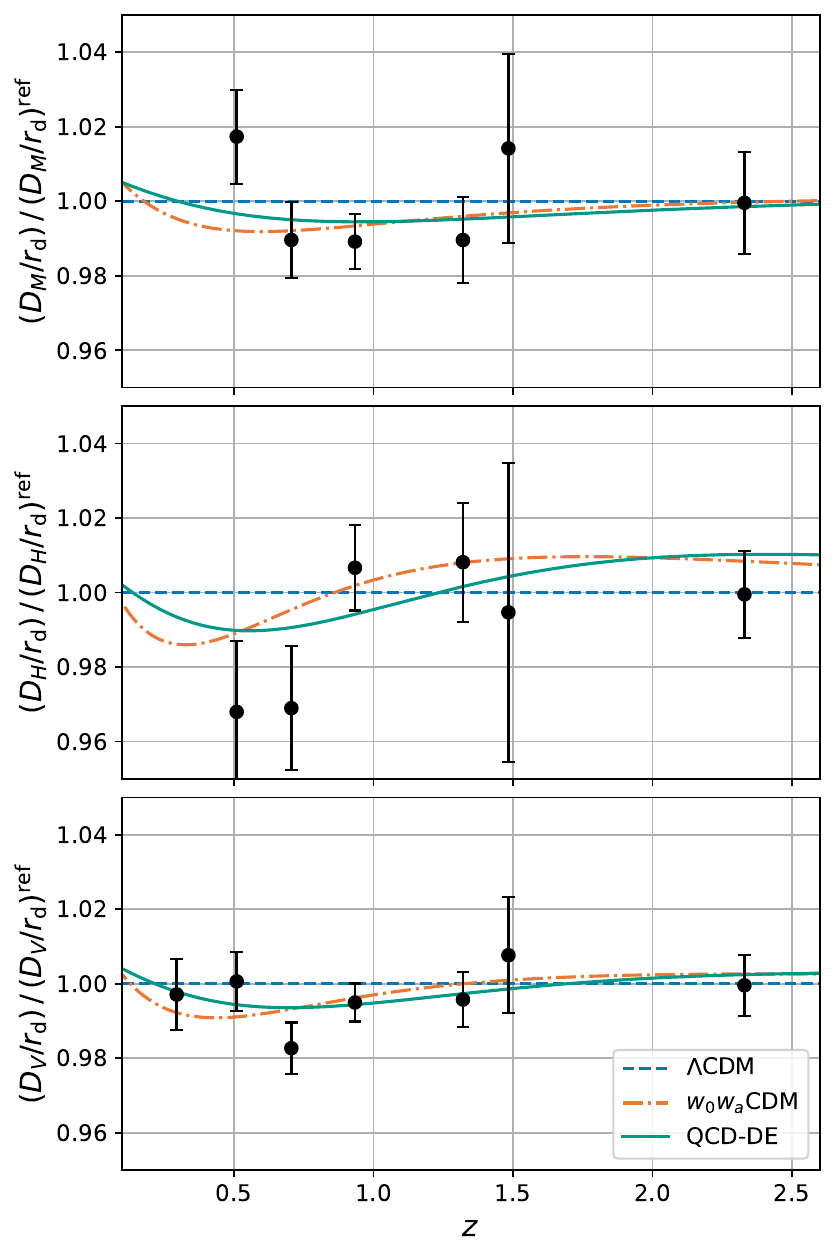}
    \caption{Residuals of the BAO distance indicators with respect to the MAP $\Lambda\mathrm{CDM}$ model for the CMB-SPA+DD+DESI dataset combination. The top, middle, and bottom panels show $(D_M/r_d)/(D_M/r_d)_{\Lambda\mathrm{CDM}}$, $(D_H/r_d)/(D_H/r_d)_{\Lambda\mathrm{CDM}}$, and $(D_V/r_d)/(D_V/r_d)_{\Lambda\mathrm{CDM}}$, respectively.}
    \label{fig:bao_residuals}
\end{figure}

The BAO residual plots provide a physical interpretation of the dark energy models and a complementary diagnostic to the statistical evidence presented in Section~\ref{subsec:statistical_evidence}. There are three BAO distance measures: the transverse distance $\theta_{\rm BAO}(z)$, the radial distance $D_H(z)$, and the isotropic distance $D_V(z)$. They are all defined in terms of the sound horizon at the baryon drag epoch, $r_d$:

\begin{equation}
r_d\equiv \int_{z_d}^\infty \frac{c_s(z)}{H(z)}{\rm d}z,
\end{equation}
where $c_s$ is the sound speed. The three BAO distances are defined as ($D_M(z)$ is the comoving line of sight distance)
\begin{equation}
\theta_{\rm BAO}\equiv \frac{r_d}{D_M(z)}, \;\;\;
D_H(z)\equiv\frac{c}{H(z)}, \;\;\;
D_V(z)\equiv \left[zD_M^2(z)D_H(z)\right]^{1/3}.
\end{equation}
The residuals are shown in Fig.~\ref{fig:bao_residuals}. Overall, the figure demonstrates that QCD-DE modifies the late-time distance-redshift relation precisely in the redshift range where DESI has the greatest constraining power, approximately $0.5\lesssim z\lesssim1.5$.
\par
The ability of the QCD-DE model to remain qualitatively similar to the $w_0w_a\mathrm{CDM}$ model,  indicates that QCD-DE is capable of reproducing the evolving dark energy signal suggested by DESI while remaining grounded in a specific physical mechanism.

The upper panel, showing $D_M/r_d$, reveals an interesting difference between the models. While the $w_0w_a\rm CDM$ model exhibits a relatively large deviation from $\Lambda\mathrm{CDM}$,
the QCD-DE model produces a more moderate shift in the transverse distance while still passing through the DESI measurements. In this sense, QCD-DE achieves a comparable fit to the data with a smaller departure from the standard cosmological model.
\par
The middle panel, showing $D_H/r_d$, highlights a more distinctive feature. Both $w_0w_a\rm CDM$ and QCD-DE display the characteristic oscillatory behaviour relative to $\Lambda\mathrm{CDM}$, with distances smaller than $\Lambda\mathrm{CDM}$ at low redshift and larger than $\Lambda\mathrm{CDM}$ at higher redshift. However, the location of the crossing differs significantly. In the $w_0w_a\rm CDM$ case, the transition occurs below $z\sim1$, whereas for QCD-DE the crossing is shifted to $z\gtrsim1$. Interestingly, the overall amplitude of the oscillation remains comparable between the models, indicating that the main difference lies in the redshift dependence of the modification rather than its magnitude.
\par
A similar behaviour is observed in the bottom panel for $D_V/r_d$. The $w_0w_a\rm CDM$ model predicts a minimum around $z\sim0.4$, while in the QCD-DE model the minimum is shifted to significantly higher redshift, around $z\sim0.7$. This shift occurs close to the region where DESI provides its strongest leverage on the expansion history and reflects the different evolution of the effective dark energy component in the QCD-DE framework.
\par
Taken together, these results show that QCD-DE produces a coherent modification of the late-time distance-redshift relation that is consistent between the radial and transverse directions. While the $w_0w_a\rm CDM$ model captures the DESI preference phenomenologically, the QCD-DE model reproduce a similar overall behaviour through a physically motivated modification of the expansion history. The improved statistical performance of QCD-DE discussed in Section~\ref{subsec:statistical_evidence} therefore corresponds to specific and physically interpretable changes in $H(z)$ and the BAO distance measures, rather than being solely a consequence of the additional parameter volume.

\section{Conclusions} \label{sec:concl}
In this work, we compared the QCD-DE model proposed by~\citealp{VanWaerbeke:2025shm} to cosmological data, including data from the CMB, BAO and supernovae measurements. Our results show a broad consistency across the different model-comparison diagnostics considered in this work. The goodness of fit of QCD-DE is comparable to that of the $w_0w_a$CDM model, while the information criteria and Bayesian evidence generally favour QCD-DE. The consistency between $\Delta\chi^2$, AIC, DIC, and $\ln{\cal B}$ indicates that the improvement in likelihood is not erased by the Occam penalty associated with the additional parameters of the model. In this sense, QCD-DE successfully captures the DESI-preferred late-time evolution while providing a more coherent and physically motivated description of the corresponding phenomenology than the standard $\Lambda\mathrm{CDM}$ and $w_0w_a\mathrm{CDM}$ models.
\par
A particularly noteworthy result is that the QCD-DE parametrization remain statistically competitive with $\Lambda$CDM for all dataset combinations considered. For early-time datasets alone and for late-time datasets alone, the Bayesian evidence generally finds QCD-DE to be statistically indistinguishable from $\Lambda$CDM. However, when early- and late-time probes are combined, QCD-DE becomes mildly to moderately preferred. This behaviour is physically appealing, since extensions to $\Lambda$CDM are not expected to be required by observations probing a single cosmological epoch, but rather by the challenge of simultaneously describing the expansion history across multiple epochs. In contrast, $w_0w_a\mathrm{CDM}$ is more strongly affected by the Bayesian penalty associated with its additional freedom and does not achieve the same level of statistical preference.

The DESI DR2 analyses have revealed increasingly strong indications for evolving dark energy, with the exact significance depending on the supernova compilation adopted. A two-parameter equation-of-state description such as CPL captures the main phenomenological trends. However, Bayesian analyses have generally found weaker evidence for departures from $\Lambda$CDM. The DES-Dovekie recalibration is particularly important in this context. For the currently most constraining dataset combination, CMB-SPA+DD+DESI, we find $\ln{\cal B}\simeq 2.0$ for QCD-DE using LHME, compared with $\ln{\cal B}\simeq 1.02$ for $w_0w_a\rm CDM$. Thus, while the Bayesian preference for $w_0w_a\rm CDM$ remains weak, the evidence for QCD-DE reaches the moderate range. This suggests that QCD-DE captures the DESI-preferred late-time evolution of DE more efficiently than the standard phenomenological CPL parametrization.

Beyond its statistical performance, one should emphasize that the QCD-DE model advocated in this work is not an ad hoc construction, unlike the majority of dynamical dark energy proposals. Instead, it originates from a specific physical mechanism associated with the non-perturbative topological structure of the QCD vacuum. The framework naturally addresses several long-standing conceptual difficulties associated with dark energy, including the coincidence problem, the drastic separation of scales, and the apparent weakness of the dark energy sector. In particular, the model provides a natural explanation for why dark energy becomes relevant only at the present epoch.

Within the phenomenological framework developed in this work, this behaviour is directly linked to the adiabatic condition of Eq.~(\ref{eq:adiabatic approximation}). This condition is strongly violated at high redshift, only marginally satisfied today, and becomes increasingly accurate in the future evolution of the Universe. Consequently, the QCD-induced vacuum contribution is naturally suppressed during radiation and matter domination, gradually becomes important at late times, and asymptotically approaches a de Sitter state as $z\rightarrow-1$, as illustrated in Figs.~(\ref{fig:adiabatic}) and (\ref{fig:w models}). The resulting dark energy scale is determined by the QCD scale $\Lambda_{\rm QCD}$, yielding an energy density of the correct order of magnitude without introducing any new fundamental scale. Remarkably, the same parameter $\Lambda_{\rm QCD}$ also determines the characteristic expansion timescale of the Universe today.

The QCD-DE model discussed in this work is a well motivated physical model for dark energy, not relying on introducing new degrees of freedom to drive the late time cosmological accelerated expansion. It offers a natural resolution to several long-standing fine-tuning problems, including the coincidence problem, the drastic separation of scales, and the unnatural weakness of interactions. Future cosmological surveys will significantly improve measurements of the late-time expansion history and provide more stringent tests of evolving dark energy. If the DESI indications persist, well motivated models such as QCD-DE may offer a promising alternative to purely phenomenological parametrizations and provide a direct connection between cosmological observations and the non-perturbative vacuum structure of QCD.
A possible direction to pursue in future work is to go beyond the phenomenological form of the switch function $\beta$ presented in this work, such as those suggested in \cite{VanWaerbeke:2025shm} inspired by 2d Schwinger model.

\section*{Acknowledgements}
We thank the organizers of the Les Houches Summer School 2025, where this work was initiated, and Julien Lesgourgues for valuable contributions and discussions. We are also grateful to Alan Heavens for helpful comments on the Bayesian evidence estimation. DHL is supported by an EPSRC studentship. CvdB is supported by the Lancaster–Sheffield Consortium for Fundamental Physics under STFC grant: ST/X000621/1. EDV is supported by a Royal Society Dorothy Hodgkin Research Fellowship. LVW and AZ are supported in part by the Natural Sciences and Engineering Research Council of Canada.  This article is based upon work from the COST Action CA21136 - ``Addressing observational tensions in cosmology with systematics and fundamental physics (CosmoVerse)'', supported by COST - ``European Cooperation in Science and Technology''. We acknowledge IT Services at The University of Sheffield for the provision of services for High Performance Computing. 

\begin{appendix}
\section{QCD-DE Constraints}\label{ap:constraints}

The $68\%$ CL constraints for QCD-DE under the combinations of datasets discussed in Sec. (\ref{subsec:datasets}) are listed in Tab. (\ref{tab:qcd constraints}).

\begin{table*}[h!]
    \renewcommand{\arraystretch}{1.3}
    \centering
    \begin{tabular}{c || c c | c c c}
    \hline\hline
    Parameters & PP+DESI & DD+DESI & CMB-SPA & CMB-SPA+PP+DESI & CMB-SPA+DD+DESI \\
    \hline
    $H_0$ & $70^{+10}_{-6}$ & $70^{+10}_{-6}$ & $62.24^{+0.93}_{-3.2}$ & $67.74\pm 0.58$ & $67.74\pm 0.52$ \\
    $\Omega_bh^2$ & $0.0264^{+0.013}_{-0.0052}$ & $0.0263^{+0.013}_{-0.0052}$ & $0.022390\pm 0.000096$ & $0.022428\pm 0.000094$ & $0.022423\pm 0.000093$ \\
    $\Omega_ch^2$ & $0.123^{+0.025}_{-0.032}$ & $0.125^{+0.026}_{-0.033}$ & $0.1206\pm 0.0010$ & $0.11955\pm 0.00078$ & $0.11957\pm 0.00078$ \\
    $\ln(10^{10}A_s)$ & $---$ & $---$ & $3.052\pm 0.011$ & $3.050\pm 0.010$ & $3.049\pm 0.010$ \\
    $n_s$ & $---$ & $---$ & $0.9683\pm 0.0035$ & $0.9710\pm 0.0030$ & $0.9710\pm 0.0031$ \\
    $\tau_\mathrm{reio}$ & $0.06$ (fixed) & $0.06$ (fixed) & $0.0552\pm 0.0057$ & $0.0553\pm 0.0054$ & $0.0552\pm 0.0055$ \\
    \hline
    $z_q$ & $0.64^{+0.29}_{-0.42}$ & $0.64^{+0.30}_{-0.37}$ & $< 1.77$ & $0.66^{+0.32}_{-0.16}$ & $0.68^{+0.31}_{-0.14}$ \\
    $\Delta z_q$ & $1.24^{+0.25}_{-0.73}$ & $1.17^{+0.23}_{-0.70}$ & $1.50^{+0.66}_{-0.95}$ & $0.83^{+0.14}_{-0.18}$ & $0.82^{+0.13}_{-0.18}$ \\
    $\overline{H}$ & $61^{+9}_{-8}$ & $61^{+9}_{-7}$ & $45.4^{+2.8}_{-8.9}$ & $59.1^{+3.8}_{-4.4}$ & $58.8^{+3.5}_{-4.3}$ \\
    $A_q$ & $1.26^{+0.10}_{-0.24}$ & $1.241^{+0.097}_{-0.23}$ & $1.23^{+0.13}_{-0.24}$ & $1.147^{+0.040}_{-0.12}$ & $1.140^{+0.035}_{-0.11}$ \\
    \hline
    $\Omega_m$ & $0.304^{+0.018}_{-0.015}$ & $0.306^{+0.019}_{-0.015}$ & $0.373^{+0.040}_{-0.016}$ & $0.3109\pm 0.0054$ & $0.3109\pm 0.0049$ \\
    $\sigma_8$ & $0.74^{+0.16}_{-0.36}$ & $0.74^{+0.17}_{-0.35}$ & $0.7795^{+0.0067}_{-0.019}$ & $0.8091\pm 0.0060$ & $0.8090\pm 0.0056$ \\
    $S_8$ & $0.75^{+0.16}_{-0.36}$ & $0.75^{+0.17}_{-0.36}$ & $0.868^{+0.029}_{-0.014}$ & $0.8235\pm 0.0064$ & $0.8235\pm 0.0063$ \\
    $r_\mathrm{drag}$ & $144.5^{+8.5}_{-22}$ & $144.3^{+8.8}_{-22}$ & $146.85\pm 0.25$ & $147.09\pm 0.21$ & $147.09\pm 0.21$ \\
    $r_\mathrm{drag}h$ & $99.49\pm 0.87$ & $99.53\pm 0.79$ & $91.4^{+1.4}_{-4.8}$ & $99.64\pm 0.85$ & $99.64\pm 0.76$ \\
    \hline\hline
    \end{tabular}
    \caption{Constraints on Cosmological parameters of the QCD-DE model across different combinations of cosmological datasets.}
    \label{tab:qcd constraints}
\end{table*}
\section{Model Comparison Results}\label{ap:Model comparison}

The Model comparison metrics of $w_0w_a\mathrm{CDM}$ and QCD-DE cosmologies in comparison to $\Lambda\mathrm{CDM}$ for the datasets: CMB-SPA, PP+DESI, DD+DESI, CMB-SPA+PP+DESI and CMB-SPA+DD+DESI are listed below in Tab. (\ref{tab:model_comparison}).

\begin{table*}[h]
\renewcommand{\arraystretch}{1.3}
\centering
\small
\setlength{\tabcolsep}{5pt}
\begin{tabular}{l | cc | cc | cc | cc}
\hline\hline
 & \multicolumn{2}{c|}{$\Delta\chi^2$} & \multicolumn{2}{c|}{$\Delta\mathrm{AIC}$} & \multicolumn{2}{c|}{$\Delta\mathrm{DIC}$} & \multicolumn{2}{c}{$\ln\mathcal{B}$} \\
Dataset & $w_0w_a$ & QCD-DE & $w_0w_a$ & QCD-DE & $w_0w_a$ & QCD-DE & $w_0w_a$ & QCD-DE \\
\hline
CMB-SPA          & -3.51  & 0.30   & 1.31  & 4.48  & -2.80  & 1.26   & $0.46\pm0.02$  & $-0.72\pm0.01$ \\
PP+DESI          & -4.89  & -6.00  & -0.89 & -2.00 & -0.74  & -0.54  & $-2.86\pm0.02$ & $0.00\pm0.02$  \\
DD+DESI          & -7.19  & -7.19  & -3.19 & -3.19 & -3.10  & -1.21  & $-1.72\pm0.03$ & $0.28\pm0.03$  \\
\hline
CMB-SPA+PP+DESI  & -11.00 & -10.84 & -4.45 & -3.49 & -8.30  & -10.37 & $-0.65\pm0.02$ & $1.72\pm0.01$  \\
CMB-SPA+DD+DESI  & -12.85 & -11.21 & -7.50 & -5.15 & -11.63 & -10.42 & $1.02\pm0.02$  & $2.02\pm0.02$  \\
\hline\hline
\end{tabular}
\caption{Model comparison statistics for $w_0w_a\mathrm{CDM}$ and QCD-DE relative to $\Lambda\mathrm{CDM}$. $\Delta\chi^2 = \chi^2_\mathrm{model}-\chi^2_{\Lambda\mathrm{CDM}}$, $\Delta\mathrm{AIC} = \mathrm{AIC}_\mathrm{model} - \mathrm{AIC}_{\Lambda\mathrm{CDM}}$, $\Delta\mathrm{DIC} = \mathrm{DIC}_\mathrm{model} - \mathrm{DIC}_{\Lambda\mathrm{CDM}}$ and $\ln\mathcal{B} = \ln\mathcal{Z}_\mathrm{model}-\ln\mathcal{Z}_{\Lambda\mathrm{CDM}}$.}
\label{tab:model_comparison}
\end{table*}
\end{appendix}

\newpage
\bibliography{biblio}
\bibliographystyle{aasjournalv7}

\end{document}